\documentclass[compsoc,10pt]{IEEEtran}
\usepackage{graphicx}
\usepackage[cmex10]{amsmath}
\usepackage[caption=false,font=footnotesize]{subfig}
\usepackage{color}
\usepackage{txfonts}
\usepackage{pgfplots}
\usetikzlibrary{arrows}
\usepackage{booktabs,array}
\usepackage{multirow}
\usepackage{url}
\usepackage{enumerate}
\usepackage{algpseudocode}
\usepackage{enumitem}
\usepackage{float}
\newcommand{\mus}{$\,\muup$s}
\renewcommand{\eqref}[1]{Eq.~(\ref{#1})}

\newfloat{algorithm}{t}{lop}
\floatname{algorithm}{Algorithm}
\newlist{scenario}{enumerate}{1}
\setlist[scenario]{label=\textit{Scenario }\Roman*:, align=left}
\newlist{rules}{enumerate}{1}
\setlist[rules]{label=\Roman*.}
\newlist{groups}{enumerate}{1}
\setlist[groups]{label=\Alph*), align=left}
\newcommand{\SC}{\mathcal{S}}
\newcommand{\Sa}{\mathcal{S}_1}
\newcommand{\Sb}{\mathcal{S}_2}
\newcommand{\N}{\mathcal{N}}

\newcommand{\Fa}{\mathcal{F}_1}
\newcommand{\Fb}{\mathcal{F}_2}
\DeclareMathOperator*{\argmin}{\arg min}
\newcommand{\rev}[1]{\textcolor{black}{#1}}
\newcounter{MYtempeqncnt}

\graphicspath{{figures/}}

\begin{document}
    
\title{RCFD: A Novel Channel Access Scheme for Full-Duplex Wireless Networks
Based on Contention in Time and Frequency Domains}

\author{Michele Luvisotto, \IEEEmembership{Student Member, IEEE}, Alireza 
Sadeghi \IEEEmembership{Student Member, IEEE},\\ Farshad Lahouti 
\IEEEmembership{Senior Member, IEEE}, Stefano Vitturi, 
\IEEEmembership{Senior Member, IEEE}, Michele Zorzi, 
\IEEEmembership{Fellow, IEEE}\IEEEcompsocitemizethanks{
\IEEEcompsocthanksitem  M. Luvisotto and M. Zorzi are with the Department of 
Information Engineering, University of Padova, Via Gradenigo 6/B, 35131 Padova,
Italy. E-mail: {luvisott, zorzi}@dei.unipd.it
\IEEEcompsocthanksitem  A. Sadeghi is with the Electrical and Computer 
Engineering Department, University of Minnesota, 200 Union Street SE, 4-174 
Keller Hall, Minneapolis, MN 55455-0170. E-mail: sadeg012@umn.edu
\IEEEcompsocthanksitem  F. Lahouti is with the Electrical Engineering 
Department, California Institute of Technology, 1200 E California Blvd, MC 
136-93, Pasadena, CA 91125. E-mail: lahouti@caltech.edu
\IEEEcompsocthanksitem  S. Vitturi is with CNR-IEIIT, National Research 
Council of Italy, Via Gradenigo 6/B, 35131 Padova, Italy. E-mail: 
stefano.vitturi@ieiit.cnr.it
}
\thanks{Part of this work was presented at the 2016 IEEE 
International Conference of Communications (ICC) \cite{Luvisotto2016}}}

\IEEEtitleabstractindextext{
\begin{abstract}
In the last years, the advancements in signal processing and integrated 
circuits technology allowed several research groups to develop working 
prototypes of in--band full--duplex wireless systems.
The introduction of such a revolutionary concept is promising in terms of 
increasing network performance, but at the same time poses several new 
challenges, especially at the MAC layer. Consequently, innovative channel 
access strategies are needed to exploit the opportunities provided by 
full--duplex while dealing with the increased complexity derived from its 
adoption.
In this direction, this paper proposes \rev{RTS/CTS in the Frequency Domain 
(RCFD),} a MAC layer scheme for full--duplex ad hoc wireless networks, based on 
the idea of time--frequency channel contention. 
According to this approach, different OFDM subcarriers are used to coordinate 
how nodes access the shared medium.
The proposed scheme leads to efficient transmission scheduling with the result 
of avoiding collisions and exploiting full--duplex opportunities. 
The considerable performance improvements with respect to standard and 
state--of--the--art MAC protocols for wireless networks are highlighted through 
both theoretical analysis and network simulations.   
\end{abstract}
}
\maketitle

\IEEEraisesectionheading{
\section{Introduction}
\label{sec:introduction}}

Innovation in Medium Access Control (MAC) plays a crucial role in the evolution of wireless networks. The purpose of MAC protocols is to efficiently coordinate the use of a shared communication 
medium by a large number of users. Depending on the network architecture, the application and the target performance, MAC protocols may be designed and operate either in a centralized or in a distributed fashion.
Currently implemented strategies are able to provide high throughput and 
acceptable fairness. However, their performance in terms of delay or efficiency 
(particularly when the payload size is small) is in general low. 
Moreover, distributed MAC schemes for wireless networks generally suffer from 
several issues, such as the hidden terminal (HT) and 
exposed terminal (ET) problems, that may result in considerable performance 
degradation \cite{Wang2012}.

Many of these issues are related to inherent limitations of wireless networks, 
one of the most important being the so--called half--duplex (HD) constraint,
i.e., the impossibility for a radio to transmit and receive in the same 
frequency band at the same time. Bidirectional communication is often 
facilitated by emulating full--duplex (FD) communication, using time--division 
duplex (TDD) or frequency--division duplex (FDD), at the expense of a reduction 
in the achievable throughput.
In the past years, however, several research groups presented working 
prototypes of FD wireless systems \cite{Choi2010,Duarte2010} that exploited 
advancements in analog circuit design and digital signal processing techniques to accomplish simultaneous transmission and reception in the same frequency band.
These results were followed by further research efforts aimed at exploiting FD 
to enhance the overall network performance.

The possibility for a node to receive and transmit at the same time increases 
the exposure to interference and considerably complicates the management of 
spatial reuse and scheduling of transmissions. Consequently, the design of new 
channel access schemes to efficiently exploit the FD capabilities and produce 
significant performance gains compared to currently deployed HD systems 
represents a very important and timely research topic and is the focus of this 
research. Assuming the use of orthogonal frequency division multiplexing (OFDM) 
as an efficient physical layer technology for communications over wideband 
wireless channels, here we specifically present a novel distributed MAC 
protocol which benefits from contention in both time and frequency domains.

In the sequel, we first briefly review the literature on full--duplex wireless MAC and contention resolution in the frequency domain and then present a summary of the contributions and the organization of this article. 

\subsection{Full--duplex wireless \rev{and time--based channel access}}
\label{subsec:related_fd}

The main challenge in achieving FD wireless communication is the 
self--interference (SI) in the receive chain when transmission and reception 
occur simultaneously. 
Indeed, due to the much shorter distance, the power of the signal emitted by a node at its own receiver is much higher than that of any other received signal.
This may prevent a successful reception when a transmission is taking place.
Hence, high performance FD wireless communication may only be feasible when self--interference is effectively canceled.

While the first experimental tests concerned with SI cancellation date back to 
1998 \cite{Chen1998}, only in 2010 did some research groups independently present the first working prototypes \cite{Choi2010,Duarte2010,Radunovic2010}.
The proposed methods exploited antenna placement, analog circuit 
design, digital domain techniques, or a combination of these approaches, with 
the aim of reducing the SI power to the noise floor level.
A very good review of the state of the art of FD wireless systems and SI 
cancellation techniques can be 
found in \cite{Sabharwal2013}. 

Several full--duplex MAC protocols for both infrastructure and ad hoc wireless 
networks have been reported in the scientific literature. A complete survey is available in \cite{Kim2015}.
In the \textit{infrastructure} configuration, some schemes have been developed 
for the case of asymmetric traffic, that aim at identifying FD opportunities 
and solving HT problems, through either busy tones \cite{Jain2011} or header 
snooping, shared backoff and virtual contention resolution \cite{Sahai2011}. 
In contrast to the centralized scheme proposed in \cite{Kim2013}, these works 
do not consider the interference between nodes. The authors in \cite{Choi2015} 
proposed a power--controlled MAC, where the transmit power of each node is 
adapted in order to maximize the signal--to--interference--plus--noise ratio of 
FD transmissions.
More strategies are available for \textit{ad hoc} networks, such as that 
mentioned in
\cite{Singh2011}, which proposes a distributed scheduling protocol aimed at 
enhancing efficiency while preserving fairness among the scheduled links. 
To cope with asymmetric traffic, the works in \cite{Duarte2014} and 
\cite{Cheng2013a} make use of Request To Send (RTS)/Clear To Send (CTS) packets to identify FD transmission opportunities. The MAC scheme proposed in \cite{Goyal2013} deals with contention resolution techniques to handle inter--node interference in FD networks. 
Other works propose solutions able to enhance the end--to--end 
performance in multi--hop FD networks, e.g., \cite{Miura2012}, where the use 
of directional antennas is addressed, \cite{Sadeghi2014}, where frequency 
reuse to enhance outage probability is investigated, and \cite{Tamaki2013}, 
that proposes synchronous channel access. Finally, cross--layer approaches have 
been proposed, in which PHY layer techniques, such as node signatures 
\cite{Srinivasan2013} and attachment coding \cite{Wang2012}, are exploited to 
schedule transmissions in FD MAC schemes.

All the presented MAC protocols adopt a time domain approach to resolve 
contentions and identify FD communication opportunities. Though widely adopted, 
this strategy generally relies on the exchange of additional control frames, 
thus decreasing the efficiency. Moreover, such class of MAC schemes often 
resort to random waiting intervals (backoffs) to avoid collisions and preserve 
fairness among users. This in turn increases the randomness in packet delivery 
and 
hampers the ability to provision quality of service (QoS) guarantees. 

\subsection{Frequency--based channel access for half--duplex}

In an attempt to overcome the limitations of standard channel access schemes 
for wireless networks in the time domain, researchers have proposed to move the 
channel contention procedure to the frequency domain \cite{Sen2010}. 
Such an approach exploits OFDM modulation at the physical layer, which provides an ordered set of subchannels or subcarriers (SCs), equally spaced in frequency within a single wideband wireless channel.
The idea is to let the nodes contend for the channel by randomly 
selecting one of these SCs and assign the channel to the node that has chosen, 
for example, the one with the lowest frequency. This resolves contention in a short deterministic time, even for a large number of nodes, compared to 
conventional time--domain schemes, such as the Carrier Sense Multiple Access 
with Collision Avoidance (CSMA/CA) protocol adopted by IEEE 802.11 networks. 
The approach was upgraded and extended to  handle multiple collision 
domains in \cite{Sen2011}, where the backoff to frequency (BACK2F) protocol was 
introduced. 
A similar strategy was suggested in \cite{Zhang2012}, where the set of 
available SCs is divided into two subsets, one destined to random 
contention and the other to node identification. Here the ACK procedure was 
also moved to the frequency domain, allowing a further improvement of the efficiency. 

Although this approach is promising in that it resolves contentions in a deterministic amount of time, it still suffers from certain issues that affect MAC in wireless ad hoc networks, such as HT and ET.
Moreover, none of the currently proposed frequency domain protocols is designed to handle channel access in FD wireless networks, while the availability of a large number of SCs in OFDM networks can be exploited to effectively identify and select FD opportunities. In addition, it has been suggested that FD communications could help limit the SC leakage problem, which affects the performance of MACs based on frequency domain contention \cite{Sen2011}.

\subsection{Contributions and Organization of this Article}

This paper proposes a MAC layer protocol for ad hoc FD wireless networks based on time--frequency contention that is capable of efficiently exploiting FD transmission opportunities and 
resolving collisions in a short and deterministic amount of time.
 
To this end, we propose a frequency domain MAC with multiple 
contention rounds in time, each using an OFDM symbol. 
This framework is exploited to advertise the transmission intentions of the 
nodes and to select, within each contention domain, the pair of nodes that will 
actually perform a data exchange. 
This strategy resembles the time--domain RTS/CTS often adopted in IEEE 802.11 
networks \cite{ieee80211std}, and therefore we refer to it as RTS/CTS in the 
Frequency Domain (RCFD). 
The presented scheme is fully distributed, effectively handles multiple 
contention domains, and preserves sufficient randomness to ensure fairness 
among different users. To the best of the authors' knowledge, this is the first 
work that combines channel access in the time and frequency domains and applies 
it to FD wireless networks.

To assess the performance of the proposed RCFD MAC protocol and compare it 
against state--of--the--art solutions, we present both theoretical analysis and 
simulation results. As a first step, we theoretically analyze the saturation 
throughput of the RCFD protocol in a network with a single 
collision domain. To provide an effective benchmark, an original theoretical 
analysis of the MAC protocols proposed in \cite{Duarte2014} and \cite{Sen2011} 
is developed. A further performance evaluation for the more general case with 
multiple collision domains is presented using network simulations. 
Again, the performance of RCFD is compared with that of state--of--the--art 
MAC protocols, that have been purposely implemented in the same simulation 
platform.

The proposed approach is able to take the best out of the two strategies previously presented, namely time--domain and frequency--domain contention. 
Indeed, compared to frequency--domain MAC protocols, such as \cite{Sen2011}, the
proposed scheme allows to eliminate the HT issue, exploiting the multiple round RTS/CTS procedure. Moreover, compared against previously reported time--domain MAC protocols for FD wireless networks, such as \cite{Duarte2014}, RCFD exhibits an increased efficiency as well as a reduced delay. 

The rest of the paper is organized as follows. The \rev{basic version of the}
RCFD MAC protocol is presented in Section~\ref{sec:protocol}, together with 
some examples of its operation. 
\rev{Section~\ref{sec:optimization} discusses the assumptions made during 
protocol design, explores its limitations and outlines some possible 
optimizations.}
Section~\ref{sec:theoretical} contains a theoretical analysis of RCFD 
and of several related wireless MAC protocols from the literature. In 
Section~\ref{sec:simulation_ns3}, the proposed protocol is evaluated using 
network simulations. Finally, Section~\ref{sec:conclusions} concludes the paper 
and outlines some future research directions.

\section{The RCFD Full--duplex MAC Protocol}
\label{sec:protocol}

The RCFD algorithm is a channel access scheme based on a time and frequency domain approach. 
According to this strategy, not only the medium contention, but also 
transmission identification and selection are performed over multiple 
consecutive frequency domain contention rounds.

\subsection{System model}
\label{subsec:assumptions}

RCFD is designed for an ad hoc wireless network composed of $N$ nodes with the 
same priority. 
Each node is assumed to have perfect FD capabilities, i.e., it can simultaneously 
receive a signal while transmitting in the same frequency band with perfect self--interference cancellation. 
OFDM is adopted at the physical layer to transmit consecutive symbols over a 
set of $S$ subcarriers. During the channel contention phase only, nodes 
transmit on single SCs while listening to the whole channel. In the data 
transmission phase, instead, only one pair of nodes transmit and receive in 
each collision domain, exploiting all SCs available in the selected channel, as 
generally done in existing IEEE 802.11 networks \cite{ieee80211std}. 

The proposed protocol relies on some assumptions that ensure its correct 
behavior. The validity of these assumptions as well as the possibility of 
relaxing them will be discussed in Section~\ref{sec:optimization}. 
We first suppose that all nodes have data to send and try to access the 
channel simultaneously. 
The communication channel is assumed ideal (no external interference, fading or 
path loss), so that each node can hear every other node within its coverage 
range. However, there can be multiple collision domains, i.e., the range of a 
node may 
not include all the nodes in the network.

We assume that a unique association between each node and two OFDM subcarriers 
is initially established at network setup, maintained fixed throughout all 
operations and available to each node. 
More specifically, defining $\SC=\{s_1,\dots,s_S\}$ as the set of 
available SCs, we split it in two non--overlapping parts $\Sa$ and $\Sb$. 
Taking $\N=\{n_1,\dots,n_N\}$ as the set of network nodes, a mapping is defined 
by the two functions
\begin{equation}
\Fa: \N \to \Sa,\quad\Fb: \N \to \Sb
\end{equation}
that uniquely link any node with an associated SC in each set.
A simple implementation of such a map can be obtained by taking 
$\Sa=\{s_1,\dots,s_{S/2}\}$, $\Sb=\{s_{S/2+1},\dots,s_S\}$ and defining 
$\Fa(n_i)=s_i,\;\Fb(n_i)=s_{i+S/2},\;i=1,\dots,N$.
It is worth stressing that the correspondence between a node and each of the 
two SCs must be unique, i.e., $\Fa(n_i)\neq\Fa(n_j)$ and $\Fb(n_i)\neq\Fb(n_j)$ 
for every $i\neq j$.
Finally, it has to be noted that the assumed mapping imposes a constraint on 
the number of nodes in the network. Indeed, since each node must be uniquely 
associated with two OFDM SCs, the total number of nodes has to be less
than or equal to $S/2$. 

\subsection{Channel contention scheme}
\label{subsec:channel_access}

The channel access procedure is composed of three consecutive contention rounds in the 
frequency domain. The first round starts after each node has sensed the channel 
and found it idle for a certain period of time $T_{scan}$. Each round consists 
in the transmission of an OFDM symbol and its duration is set to 
$T_{round}=T_{sym}+2T_p$ to accommodate for signal propagation, which takes a 
time $T_p$ each way \cite{Sen2011}. Therefore, the access procedure 
takes a fixed time of 
\begin{equation}
\label{eq:channel_access}
T_{acc}=T_{scan}+3 T_{round}
\end{equation}
As an example, if an IEEE 802.11g network is considered, standard values for 
these parameters are $T_{scan}=$ 28~\mus\ (the duration of a DCF inter--frame 
space interval), $T_{sym}=$ 4~\mus, and $T_p=$ 1~\mus, thus obtaining 
$T_{acc}=$ 46~\mus.

In the following, we outline the steps performed by every node in each 
contention round.

\subsubsection{First round - randomized contention}

Every node that has data to send and has found the channel idle for a 
$T_{scan}$ period, randomly selects an SC from the whole set $\SC$ and 
transmits a symbol only on that SC, while listening to the whole channel band. 

We denote with $\bar{s}_i$ the SC chosen by node $n_i$ and we also indicate 
with $\SC_i^1$ the set of SCs that actually carried a symbol during the first 
contention round, as perceived by node $n_i$. 

Node $n_i$ is defined as \textit{primary transmitter} (PT) if and only if the 
following condition holds
\begin{equation}\label{eq:pt}
\bar{s}_i=\min\limits_j\left[s_j\in\SC_i^1\right]
\end{equation}
i.e., the lowest--frequency SC among those carrying data is the one chosen by 
the node itself. 
It is noteworthy that, in a realistic scenario with multiple collision 
domains, several nodes in the network can be selected as PTs.
Moreover, if multiple nodes in the same collision domain pick the same 
lowest--frequency SC, they are all selected as PTs. This potential collision 
will be resolved in the following contention rounds\rev{, as explained in 
Section~\ref{subsec:fading}}.

\begin{figure*}[t!]
\begin{center}
\scalebox{0.6}{
\begin{tikzpicture}
	\node at (0,-3) {Node $n_1$};
	\draw[draw=black,very thick] (1,-3)--(7,-3);
	\draw (1.5,-3) -- (1.5,-3.2);
	\node at (1.5,-3.4) {$s_1$};
	\draw (2.5,-3) -- (2.5,-3.2);
	\node at (2.5,-3.4) {$s_2$};
	\draw (3.5,-3) -- (3.5,-3.2);
	\node at (3.5,-3.4) {$s_3$};
	\draw (4.5,-3) -- (4.5,-3.2);
	\node at (4.5,-3.4) {$s_4$};
	\draw (5.5,-3) -- (5.5,-3.2);
	\node at (5.5,-3.4) {$s_5$};
	\draw (6.5,-3) -- (6.5,-3.2);
	\node at (6.5,-3.4) {$s_6$};	
	\draw[-latex, ultra thick,draw=green,fill=green] (4.5,-3) -- (4.5,-2);
	\node at (0,-4.5) {Node $n_2$};
	\draw[draw=black,very thick] (1,-4.5)--(7,-4.5);
	\draw (1.5,-4.5) -- (1.5,-4.7);
	\node at (1.5,-4.9) {$s_1$};
	\draw (2.5,-4.5) -- (2.5,-4.7);
	\node at (2.5,-4.9) {$s_2$};
	\draw (3.5,-4.5) -- (3.5,-4.7);
	\node at (3.5,-4.9) {$s_3$};
	\draw (4.5,-4.5) -- (4.5,-4.7);
	\node at (4.5,-4.9) {$s_4$};
	\draw (5.5,-4.5) -- (5.5,-4.7);
	\node at (5.5,-4.9) {$s_5$};
	\draw (6.5,-4.5) -- (6.5,-4.7);
	\node at (6.5,-4.9) {$s_6$};	
	\draw[-latex, ultra thick,draw=red,fill=red] (4.5,-4.5) -- (4.5,-4);
	\draw[-latex, ultra thick,draw=red,fill=red] (5.5,-4.5) -- (5.5,-4);
	\node at (0,-6) {Node $n_3$};
	\draw[draw=black,very thick] (1,-6)--(7,-6);
	\draw (1.5,-6) -- (1.5,-6.2);
	\node at (1.5,-6.4) {$s_1$};
	\draw (2.5,-6) -- (2.5,-6.2);
	\node at (2.5,-6.4) {$s_2$};
	\draw (3.5,-6) -- (3.5,-6.2);
	\node at (3.5,-6.4) {$s_3$};
	\draw (4.5,-6) -- (4.5,-6.2);
	\node at (4.5,-6.4) {$s_4$};
	\draw (5.5,-6) -- (5.5,-6.2);
	\node at (5.5,-6.4) {$s_5$};
	\draw (6.5,-6) -- (6.5,-6.2);
	\node at (6.5,-6.4) {$s_6$};
	\draw[-latex, ultra thick,draw=green,fill=green] (5.5,-6) -- (5.5,-5);
	\node at (4,-7) {First round};
	\draw[draw=black,very thick] (8,-3)--(14,-3);
	\draw (8.5,-3) -- (8.5,-3.2);
	\node at (8.5,-3.4) {$s_1$};
	\draw (9.5,-3) -- (9.5,-3.2);
	\node at (9.5,-3.4) {$s_2$};
	\draw (10.5,-3) -- (10.5,-3.2);
	\node at (10.5,-3.4) {$s_3$};
	\draw (11.5,-3) -- (11.5,-3.2);
	\node at (11.5,-3.4) {$s_4$};
	\draw (12.5,-3) -- (12.5,-3.2);
	\node at (12.5,-3.4) {$s_5$};
	\draw (13.5,-3) -- (13.5,-3.2);
	\node at (13.5,-3.4) {$s_6$};	
	\draw[-latex, ultra thick,draw=green,fill=green] (8.5,-3) -- (8.5,-2);
	\draw[-latex, ultra thick,draw=green,fill=green] (12.5,-3) -- (12.5,-2);
	\draw[draw=black,very thick] (8,-4.5)--(14,-4.5);
	\draw (8.5,-4.5) -- (8.5,-4.7);
	\node at (8.5,-4.9) {$s_1$};
	\draw (9.5,-4.5) -- (9.5,-4.7);
	\node at (9.5,-4.9) {$s_2$};
	\draw (10.5,-4.5) -- (10.5,-4.7);
	\node at (10.5,-4.9) {$s_3$};
	\draw (11.5,-4.5) -- (11.5,-4.7);
	\node at (11.5,-4.9) {$s_4$};
	\draw (12.5,-4.5) -- (12.5,-4.7);
	\node at (12.5,-4.9) {$s_5$};
	\draw (13.5,-4.5) -- (13.5,-4.7);
	\node at (13.5,-4.9) {$s_6$};
	\draw[-latex, ultra thick,draw=red,fill=red] (8.5,-4.5) -- (8.5,-4);
	\draw[-latex, ultra thick,draw=red,fill=red] (10.5,-4.5) -- (10.5,-4);	
	\draw[-latex, ultra thick,draw=red,fill=red] (12.5,-4.5) -- (12.5,-4);	
	\draw[draw=black,very thick] (8,-6)--(14,-6);
	\draw (8.5,-6) -- (8.5,-6.2);
	\node at (8.5,-6.4) {$s_1$};
	\draw (9.5,-6) -- (9.5,-6.2);
	\node at (9.5,-6.4) {$s_2$};
	\draw (10.5,-6) -- (10.5,-6.2);
	\node at (10.5,-6.4) {$s_3$};
	\draw (11.5,-6) -- (11.5,-6.2);
	\node at (11.5,-6.4) {$s_4$};
	\draw (12.5,-6) -- (12.5,-6.2);
	\node at (12.5,-6.4) {$s_5$};
	\draw (13.5,-6) -- (13.5,-6.2);
	\node at (13.5,-6.4) {$s_6$};
	\draw[-latex, ultra thick,draw=green,fill=green] (10.5,-6) -- (10.5,-5);
	\draw[-latex, ultra thick,draw=green,fill=green] (12.5,-6) -- (12.5,-5);
	\node at (11,-7) {Second round};
	\draw[draw=black,very thick] (15,-3)--(21,-3);
	\draw (15.5,-3) -- (15.5,-3.2);
	\node at (15.5,-3.4) {$s_1$};
	\draw (16.5,-3) -- (16.5,-3.2);
	\node at (16.5,-3.4) {$s_2$};
	\draw (17.5,-3) -- (17.5,-3.2);
	\node at (17.5,-3.4) {$s_3$};
	\draw (18.5,-3) -- (18.5,-3.2);
	\node at (18.5,-3.4) {$s_4$};
	\draw (19.5,-3) -- (19.5,-3.2);
	\node at (19.5,-3.4) {$s_5$};
	\draw (20.5,-3) -- (20.5,-3.2);
	\node at (20.5,-3.4) {$s_6$};
	\draw[-latex, ultra thick,draw=red,fill=red] (16.5,-3) -- (16.5,-2.5);	
	\draw[-latex, ultra thick,draw=red,fill=red] (18.5,-3) -- (18.5,-2.5);	
	\draw[draw=black,very thick] (15,-4.5)--(21,-4.5);
	\draw (15.5,-4.5) -- (15.5,-4.7);
	\node at (15.5,-4.9) {$s_1$};
	\draw (16.5,-4.5) -- (16.5,-4.7);
	\node at (16.5,-4.9) {$s_2$};
	\draw (17.5,-4.5) -- (17.5,-4.7);
	\node at (17.5,-4.9) {$s_3$};
	\draw (18.5,-4.5) -- (18.5,-4.7);
	\node at (18.5,-4.9) {$s_4$};
	\draw (19.5,-4.5) -- (19.5,-4.7);
	\node at (19.5,-4.9) {$s_5$};
	\draw (20.5,-4.5) -- (20.5,-4.7);
	\node at (20.5,-4.9) {$s_6$};	
	\draw[-latex, ultra thick,draw=green,fill=green] (16.5,-4.5) -- 
	(16.5,-3.5);	
	\draw[-latex, ultra thick,draw=green,fill=green] (18.5,-4.5) -- (18.5,-3.5);
	\draw[draw=black,very thick] (15,-6)--(21,-6);
	\draw (15.5,-6) -- (15.5,-6.2);
	\node at (15.5,-6.4) {$s_1$};
	\draw (16.5,-6) -- (16.5,-6.2);
	\node at (16.5,-6.4) {$s_2$};
	\draw (17.5,-6) -- (17.5,-6.2);
	\node at (17.5,-6.4) {$s_3$};
	\draw (18.5,-6) -- (18.5,-6.2);
	\node at (18.5,-6.4) {$s_4$};
	\draw (19.5,-6) -- (19.5,-6.2);
	\node at (19.5,-6.4) {$s_5$};
	\draw (20.5,-6) -- (20.5,-6.2);
	\node at (20.5,-6.4) {$s_6$};
	\draw[-latex, ultra thick,draw=red,fill=red] (16.5,-6) -- (16.5,-5.5);	
	\draw[-latex, ultra thick,draw=red,fill=red] (18.5,-6) -- (18.5,-5.5);	
	\node at (18,-7) {Third round};
	\node at (23,-3) {\textcolor{green}{TX subcarriers}};
	\node at (23.25,-3.75) {\textcolor{red}{Heard subcarriers}};	
\end{tikzpicture}}
\end{center}
\caption{Outcomes of contention rounds for example scenario 1.}
\label{fig:sc_scenario1}
\end{figure*}
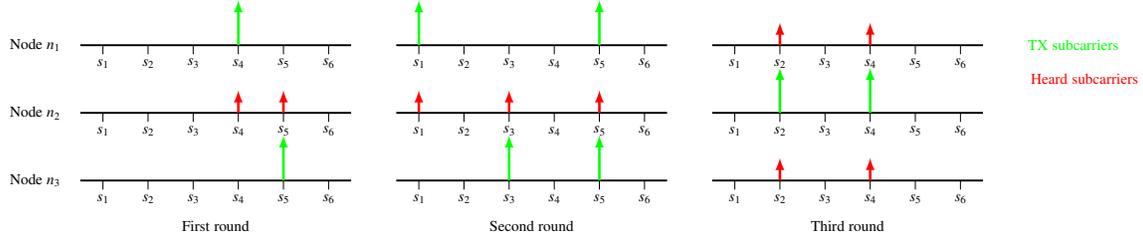

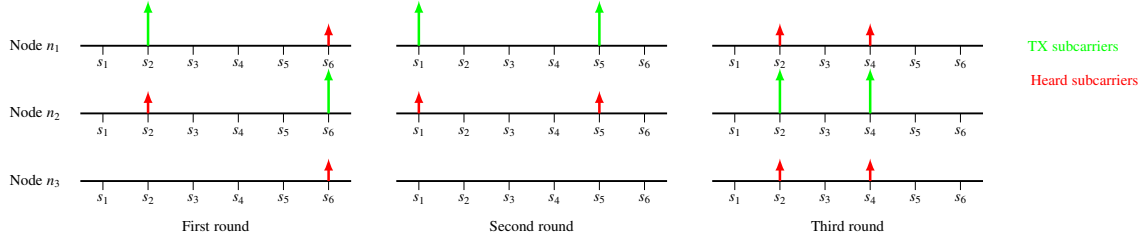
\begin{figure*}[t!]
\begin{center}
\scalebox{0.6}{
\begin{tikzpicture}
	\node at (0,-3) {Node $n_1$};
	\draw[draw=black,very thick] (1,-3)--(7,-3);
	\draw (1.5,-3) -- (1.5,-3.2);
	\node at (1.5,-3.4) {$s_1$};
	\draw (2.5,-3) -- (2.5,-3.2);
	\node at (2.5,-3.4) {$s_2$};
	\draw (3.5,-3) -- (3.5,-3.2);
	\node at (3.5,-3.4) {$s_3$};
	\draw (4.5,-3) -- (4.5,-3.2);
	\node at (4.5,-3.4) {$s_4$};
	\draw (5.5,-3) -- (5.5,-3.2);
	\node at (5.5,-3.4) {$s_5$};
	\draw (6.5,-3) -- (6.5,-3.2);
	\node at (6.5,-3.4) {$s_6$};	
	\draw[-latex, ultra thick,draw=green,fill=green] (2.5,-3) -- (2.5,-2);
	\draw[-latex, ultra thick,draw=red,fill=red] (6.5,-3) -- (6.5,-2.5);
	\node at (0,-4.5) {Node $n_2$};
	\draw[draw=black,very thick] (1,-4.5)--(7,-4.5);
	\draw (1.5,-4.5) -- (1.5,-4.7);
	\node at (1.5,-4.9) {$s_1$};
	\draw (2.5,-4.5) -- (2.5,-4.7);
	\node at (2.5,-4.9) {$s_2$};
	\draw (3.5,-4.5) -- (3.5,-4.7);
	\node at (3.5,-4.9) {$s_3$};
	\draw (4.5,-4.5) -- (4.5,-4.7);
	\node at (4.5,-4.9) {$s_4$};
	\draw (5.5,-4.5) -- (5.5,-4.7);
	\node at (5.5,-4.9) {$s_5$};
	\draw (6.5,-4.5) -- (6.5,-4.7);
	\node at (6.5,-4.9) {$s_6$};	
	\draw[-latex, ultra thick,draw=red,fill=red] (2.5,-4.5) -- (2.5,-4);
	\draw[-latex, ultra thick,draw=green,fill=green] (6.5,-4.5) -- (6.5,-3.5);
	\node at (0,-6) {Node $n_3$};
	\draw[draw=black,very thick] (1,-6)--(7,-6);
	\draw (1.5,-6) -- (1.5,-6.2);
	\node at (1.5,-6.4) {$s_1$};
	\draw (2.5,-6) -- (2.5,-6.2);
	\node at (2.5,-6.4) {$s_2$};
	\draw (3.5,-6) -- (3.5,-6.2);
	\node at (3.5,-6.4) {$s_3$};
	\draw (4.5,-6) -- (4.5,-6.2);
	\node at (4.5,-6.4) {$s_4$};
	\draw (5.5,-6) -- (5.5,-6.2);
	\node at (5.5,-6.4) {$s_5$};
	\draw (6.5,-6) -- (6.5,-6.2);
	\node at (6.5,-6.4) {$s_6$};
	\draw[-latex, ultra thick,draw=red,fill=red] (6.5,-6) -- (6.5,-5.5);
	\node at (4,-7) {First round};
	\draw[draw=black,very thick] (8,-3)--(14,-3);
	\draw (8.5,-3) -- (8.5,-3.2);
	\node at (8.5,-3.4) {$s_1$};
	\draw (9.5,-3) -- (9.5,-3.2);
	\node at (9.5,-3.4) {$s_2$};
	\draw (10.5,-3) -- (10.5,-3.2);
	\node at (10.5,-3.4) {$s_3$};
	\draw (11.5,-3) -- (11.5,-3.2);
	\node at (11.5,-3.4) {$s_4$};
	\draw (12.5,-3) -- (12.5,-3.2);
	\node at (12.5,-3.4) {$s_5$};
	\draw (13.5,-3) -- (13.5,-3.2);
	\node at (13.5,-3.4) {$s_6$};	
	\draw[-latex, ultra thick,draw=green,fill=green] (8.5,-3) -- (8.5,-2);
	\draw[-latex, ultra thick,draw=green,fill=green] (12.5,-3) -- (12.5,-2);
	\draw[draw=black,very thick] (8,-4.5)--(14,-4.5);
	\draw (8.5,-4.5) -- (8.5,-4.7);
	\node at (8.5,-4.9) {$s_1$};
	\draw (9.5,-4.5) -- (9.5,-4.7);
	\node at (9.5,-4.9) {$s_2$};
	\draw (10.5,-4.5) -- (10.5,-4.7);
	\node at (10.5,-4.9) {$s_3$};
	\draw (11.5,-4.5) -- (11.5,-4.7);
	\node at (11.5,-4.9) {$s_4$};
	\draw (12.5,-4.5) -- (12.5,-4.7);
	\node at (12.5,-4.9) {$s_5$};
	\draw (13.5,-4.5) -- (13.5,-4.7);
	\node at (13.5,-4.9) {$s_6$};
	\draw[-latex, ultra thick,draw=red,fill=red] (8.5,-4.5) -- (8.5,-4);
	\draw[-latex, ultra thick,draw=red,fill=red] (12.5,-4.5) -- (12.5,-4);	
	\draw[draw=black,very thick] (8,-6)--(14,-6);
	\draw (8.5,-6) -- (8.5,-6.2);
	\node at (8.5,-6.4) {$s_1$};
	\draw (9.5,-6) -- (9.5,-6.2);
	\node at (9.5,-6.4) {$s_2$};
	\draw (10.5,-6) -- (10.5,-6.2);
	\node at (10.5,-6.4) {$s_3$};
	\draw (11.5,-6) -- (11.5,-6.2);
	\node at (11.5,-6.4) {$s_4$};
	\draw (12.5,-6) -- (12.5,-6.2);
	\node at (12.5,-6.4) {$s_5$};
	\draw (13.5,-6) -- (13.5,-6.2);
	\node at (13.5,-6.4) {$s_6$};
	\node at (11,-7) {Second round};
	\draw[draw=black,very thick] (15,-3)--(21,-3);
	\draw (15.5,-3) -- (15.5,-3.2);
	\node at (15.5,-3.4) {$s_1$};
	\draw (16.5,-3) -- (16.5,-3.2);
	\node at (16.5,-3.4) {$s_2$};
	\draw (17.5,-3) -- (17.5,-3.2);
	\node at (17.5,-3.4) {$s_3$};
	\draw (18.5,-3) -- (18.5,-3.2);
	\node at (18.5,-3.4) {$s_4$};
	\draw (19.5,-3) -- (19.5,-3.2);
	\node at (19.5,-3.4) {$s_5$};
	\draw (20.5,-3) -- (20.5,-3.2);
	\node at (20.5,-3.4) {$s_6$};
	\draw[-latex, ultra thick,draw=red,fill=red] (16.5,-3) -- (16.5,-2.5);	
	\draw[-latex, ultra thick,draw=red,fill=red] (18.5,-3) -- (18.5,-2.5);	
	\draw[draw=black,very thick] (15,-4.5)--(21,-4.5);
	\draw (15.5,-4.5) -- (15.5,-4.7);
	\node at (15.5,-4.9) {$s_1$};
	\draw (16.5,-4.5) -- (16.5,-4.7);
	\node at (16.5,-4.9) {$s_2$};
	\draw (17.5,-4.5) -- (17.5,-4.7);
	\node at (17.5,-4.9) {$s_3$};
	\draw (18.5,-4.5) -- (18.5,-4.7);
	\node at (18.5,-4.9) {$s_4$};
	\draw (19.5,-4.5) -- (19.5,-4.7);
	\node at (19.5,-4.9) {$s_5$};
	\draw (20.5,-4.5) -- (20.5,-4.7);
	\node at (20.5,-4.9) {$s_6$};	
	\draw[-latex, ultra thick,draw=green,fill=green] (16.5,-4.5) -- 
	(16.5,-3.5);	
	\draw[-latex, ultra thick,draw=green,fill=green] (18.5,-4.5) -- (18.5,-3.5);
	\draw[draw=black,very thick] (15,-6)--(21,-6);
	\draw (15.5,-6) -- (15.5,-6.2);
	\node at (15.5,-6.4) {$s_1$};
	\draw (16.5,-6) -- (16.5,-6.2);
	\node at (16.5,-6.4) {$s_2$};
	\draw (17.5,-6) -- (17.5,-6.2);
	\node at (17.5,-6.4) {$s_3$};
	\draw (18.5,-6) -- (18.5,-6.2);
	\node at (18.5,-6.4) {$s_4$};
	\draw (19.5,-6) -- (19.5,-6.2);
	\node at (19.5,-6.4) {$s_5$};
	\draw (20.5,-6) -- (20.5,-6.2);
	\node at (20.5,-6.4) {$s_6$};
	\draw[-latex, ultra thick,draw=red,fill=red] (16.5,-6) -- (16.5,-5.5);	
	\draw[-latex, ultra thick,draw=red,fill=red] (18.5,-6) -- (18.5,-5.5);	
	\node at (18,-7) {Third round};
	\node at (23,-3) {\textcolor{green}{TX subcarriers}};
	\node at (23.25,-3.75) {\textcolor{red}{Heard subcarriers}};	
\end{tikzpicture}}
\end{center}
\caption{Outcomes of contention rounds for example scenario 2.}
\label{fig:sc_scenario2}
\end{figure*}

\subsubsection{Second round - transmission advertisement (RTS)}

Only the nodes who identify themselves as PTs during the first round transmit during the 
second round. A PT node $n_i$ that has data to send to node $n_j$ transmits a 
symbol on two SCs, namely $s_a=\Fa(n_i)\in\Sa$ and $s_b=\Fb(n_j)\in\Sb$. 
In this way, $n_i$ informs its neighbors that it is a PT and has a packet for 
$n_j$. 
This round is the so--called RTS part of the algorithm, as it resembles the 
time domain RTS procedure defined in the IEEE 802.11 standard 
\cite{ieee80211std}.
During the second round, all the nodes in the network (including the PTs) 
listen to the whole band. 
We denote as $\SC_{h,1}^2\subseteq\Sa$ and $\SC_{h,2}^2\subseteq\Sb$ the sets 
of SCs that carried a symbol during the second contention round, as observed 
by a generic node $n_h$. 

Node $n_h$ is defined as \textit{RTS receiver} (RR) if and only if the 
following condition holds
\begin{equation}
\label{eq:rr}
{\color{black}n_h\text{ is not PT}\;\wedge\;}\Fb(n_h)\in\SC_{h,2}^2
\end{equation}
i.e., at least one PT node advertised, during the second round, that it has a 
packet for $n_h$. There can be multiple RRs in the network, but a node cannot 
be both PT and RR at the same time. \rev{Indeed, \rev{according to 
\eqref{eq:rr},} even if a node that is PT 
receives an RTS (e.g., due to a first round collision where two nodes in the 
same domain selected the same lowest--frequency subcarrier), it does not take 
it into account and does not define itself as RR.}

\subsubsection{Third round - transmission authorization (CTS)}

Only the nodes selected as RRs during the second round transmit in the 
third one. Any RR node $n_h$ will select its CTS recipient as
\begin{equation}
n_l = 
\argmin\limits_{n_i}\left[\Fa(n_i)\;:\;\Fa\left(n_i\right)\in\SC_{h,1}^2\right]
\end{equation}
i.e., among the nodes that have sent an RTS to $n_h$, the one with the lowest 
corresponding SC is selected.\footnote{\rev{This choice may impair the fairness 
of the RCFD protocol if the subcarrier mapping is static. To avoid 
such a problem, periodic permutations of the maps $\Fa$ and $\Fb$ according to 
a common pseudorandom sequence can be scheduled. The exchange of broadcast 
messages advertising the new maps after each permutation might be needed to 
avoid synchronization issues among the nodes. This expedient is not implemented 
in the simulations proposed in this paper, which, however, show a good fairness 
level, as highlighted in Section~\ref{subsec:simulation_results_random}.}} 
Node $n_h$ then transmits a symbol on two SCs, namely $s_c=\Fa(n_h)\in\Sa$ and 
$s_d=\Fb(n_l)\in\Sb$. 
In this way, $n_h$ informs $n_l$ that its transmission is authorized. 
Since this round mimics the operation of the time domain CTS procedure, it 
is referred to as the CTS part of the RCFD algorithm.
During the third round, all the nodes in the network (including the RRs) 
listen to the whole channel band. We denote as $\SC_{i,1}^3\subseteq\Sa$ and 
$\SC_{i,2}^3\subseteq\Sb$ the sets of SCs that carried a symbol during the 
third round, as observed by a generic node $n_i$.

At the end of the third round, each node that has data to send needs to decide 
whether to transmit or not, according to the information gathered in the three 
rounds. 
Specifically, for a generic node $n_i$ which has a packet for node $n_j$, 
three cases can be distinguished:
\begin{rules}
\item \textit{Node $n_i$ is a PT:}\\
It transmits if and only if both these conditions are verified
\begin{align}
\label{eq:transm_pt}
\begin{split}
&\Fa(n_j)\in\SC_{i,1}^3\\
&\SC_{i,2}^3=\left\{\Fb(n_i)\right\}
\end{split}
\end{align}
i.e., the intended receiver (node $n_j$) has sent a CTS and this
is the only CTS within the contention domain of node $n_i$.

\item \textit{Node $n_i$ is an RR:}\\
It transmits (while receiving from the PT, thus enabling FD) if and only if both these 
conditions are verified
\begin{equation}
\label{eq:transm_rr}
\begin{split}
&\SC_{i,1}^2=\left\{\Fa(n_j)\right\}\\
&\SC_{i,1}^3=\left\{\Fa(n_i)\right\}
\end{split}
\end{equation}
i.e., only the intended receiver (node $n_j$) has sent an RTS and no other  
node has sent a CTS (except node $n_i$ itself).

\item \textit{Node $n_i$ is neither a PT nor an RR:}\\
It does not transmit.
\end{rules} 

We point out that not only may the nodes selected as PTs during the
first round be granted access to the channels, but also an RR can transmit, 
if the conditions in case II are verified.
This possibility is the key to enable FD transmission: a node
that has a packet for another node from which it has received an RTS can send 
it together with the primary transmission (provided that no other CTSs from
surrounding nodes were received).

\rev{We remark that the RCFD protocol only allows for \textit{bidirectional FD} 
and does not take into account the asymmetric FD 
opportunities, where a node $n_j$ receives from a node $n_i$ and transmits to a 
third node $n_k$. A modification of RCFD to accommodate for asymmetric FD is 
left for future research.}

\subsection{Examples of operation}

In order to better understand how the proposed MAC strategy works, we provide 
here two examples, for a simplified system with $N=3$ nodes and 
$S=6$ OFDM subcarriers. The simplest scheme is adopted for SC mapping, i.e.,  
$\Sa=\{s_1,s_2,s_3\}$, $\Sb=\{s_4,s_5,s_6\}$, 
$\Fa(n_i)=s_i$, $\Fb(n_i)=s_{i+3}$, $i=1,2,3$. 

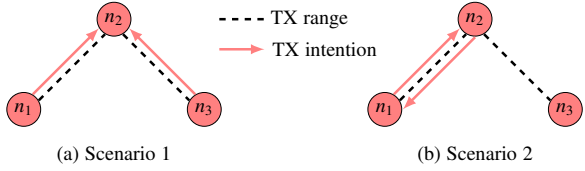
\begin{figure}[t!]
\begin{center}
\scalebox{0.8}{
\begin{tikzpicture}
	\draw[fill=red!50!white,draw=black] (0,-1) circle (8pt);
	\draw[fill=red!50!white,draw=black] (3,-1) circle (8pt);
	\draw[fill=red!50!white,draw=black] (1.5,0.5) circle (8pt);
	\draw[dashed,very thick] (0.25,-.85) -- 
	(1.35,0.25);
	\draw[-latex,draw=red!50!white,fill=red!50!white,very thick] (0.15,-.75) -- 
	(1.25,0.35);
	\draw[dashed,very thick] (1.65,0.25) -- (2.75,-.85);
	\draw[-latex,draw=red!50!white,fill=red!50!white,very thick] (2.85,-.75) -- 
		(1.75,0.35);
	\node at (0,-1) {$n_1$};
	\node at (1.5,0.5) {$n_2$};
	\node at (3,-1) {$n_3$};
	\node at (1.5,-1.75) {(a) Scenario 1};
	\draw[fill=red!50!white,draw=black] (6,-1) circle (8pt);
	\draw[fill=red!50!white,draw=black] (9,-1) circle (8pt);
	\draw[fill=red!50!white,draw=black] (7.5,0.5) circle (8pt);
	\draw[dashed,very thick] (6.25,-.85) -- 
	(7.35,0.25);
	\draw[-latex,draw=red!50!white,fill=red!50!white,very thick] (6.15,-.75) -- 
	(7.25,0.35);
	\draw[latex-,draw=red!50!white,fill=red!50!white,very thick] (6.3,-1) -- 
	(7.5,0.2);
	\draw[dashed,very thick] (7.65,0.25) -- (8.75,-.85);
	\node at (6,-1) {$n_1$};
	\node at (7.5,0.5) {$n_2$};
	\node at (9,-1) {$n_3$};
	\node at (7.5,-1.75) {(b) Scenario 2};
	\draw [dashed, very thick] (3.25,0.5) -- (4,0.5);
	\draw [-latex, draw=red!50!white,fill=red!50!white,very thick] (3.25,0) -- 
	(4,0);
	\node at (4.75,0.5) [align=left] {TX range};
	\node at (5,0) [align=left] {TX intention};
\end{tikzpicture}}
\end{center}
\caption{Topology and transmission intentions for the operation examples.}
\label{fig:setup_example}
\end{figure}

Two different example scenarios are considered. Fig.~\ref{fig:sc_scenario1} and 
Fig.~\ref{fig:sc_scenario2} show the contention rounds for scenarios 1 and 2, 
respectively, while Fig.~\ref{fig:setup_example} reports the network topology 
and the transmission intentions. 
In both scenarios, node $n_2$ is within the transmission range of nodes $n_1$ 
and $n_3$ that, however, cannot sense each other (two collision domains). 
In the first scenario, nodes $n_1$ and $n_3$ both intend to send a packet to 
$n_2$, resembling a typical HT situation. In the second one, nodes $n_1$ and 
$n_2$ have a packet for each other, representing a potential FD 
communication instance.

As seen in Fig.~\ref{fig:sc_scenario1} for scenario 1,
in the first round the two nodes with data to send randomly select two SCs as 
$\bar{s}_1=s_4$ and $\bar{s}_3=s_5$, with the result that both 
$n_1$ and $n_3$ are selected as PTs, since they cannot sense each other's 
transmissions. 
Consequently, in the second round they both transmit, causing $n_2$ to hear 
signals on SCs $s_1$, $s_3$ and $s_5$. According to \eqref{eq:rr}, $n_2$ is 
selected as RR and transmits, during the third round, on SCs $s_2$ and $s_4$. 
Finally, according to \eqref{eq:transm_pt}, node $n_1$ is allowed to transmit, 
whereas the transmission by node $n_3$ is denied, since 
$\SC_{3,2}^3=\{s_4\}$ and $\Fb(n_3)=s_6$. 
It can hence be observed that the HT problem has been identified and solved 
thanks to the RCFD strategy.

In scenario 2, as depicted in Fig.~\ref{fig:sc_scenario2}, nodes $n_1$ 
and $n_2$ participate in the first contention round, randomly selecting 
$\bar{s}_1=s_2$ and $\bar{s}_2=s_6$, therefore only $n_1$ is selected as PT.
In the second round, $n_1$ transmits on SCs $s_1$ and $s_5$, thus node $n_2$ is 
selected as RR. Finally, in the third round $n_2$ transmits on SCs $s_2$ and 
$s_4$, providing a CTS to node $n_1$. 
Since the conditions in \eqref{eq:transm_pt} are verified for $n_1$ and those 
in \eqref{eq:transm_rr} are fulfilled for $n_2$, both nodes are cleared to 
transmit, thus enabling full--duplex transmission. 
We note that if node $n_2$ had been selected as PT in the first round, the 
final outcome would have been the same ($n_1$ selected as RR and subsequently 
cleared to transmit).

\section{Protocol optimization and discussion}
\label{sec:optimization}

In this section, we discuss the assumptions on which the RCFD 
strategy is based, explore its limitations and propose some possible 
enhancements. 

\subsection{\rev{Enhancements to the subcarrier mapping scheme}}
\label{subsec:mapping}

As mentioned in Section~\ref{subsec:assumptions}, the subcarrier 
mapping upon which the RCFD scheme relies imposes a limit on the number of 
nodes in the network, which has to be no higher than $S/2$. 

It is worth stressing that the trend in wireless networks based on the IEEE 
802.11 standard is to use wider channels, that offer an ever increasing 
number of SCs. As an example, IEEE 802.11ac introduces \rev{160}~MHz 
channels, that can accommodate \rev{512} SCs and hence allow RCFD to reach up 
to \rev{256} users \cite{perahia2013next}. 

The number of nodes can be further increased even maintaining a fixed number of 
SCs if we plan to exploit the information carried in each SC. In the presented 
version of the algorithm only the presence or 
absence of data on an SC was taken into account. 
A more refined version would discriminate between the actual content of the 
symbol transmitted in a specific SC, to be able to host multiple nodes within 
the same subcarriers. 
Each SC can carry $\log_2 m$ bits if an $m$-ary modulation is adopted and, in 
this way, the maximum number of users in the system can be increased to $m\cdot 
S/2$. 
As an example, if $S=64$ SCs are available and a 64--QAM modulation is 
employed, a total of 2048 users can be hosted in the network.

Tab.~\ref{tab:example_sc} provides an example of extended subcarrier mapping in 
a system 
with $S=4$ SCs which adopts a modulation of order $m=4$, hence allowing the 
presence of 8 users. In this scenario, for instance, if node $n_1$ has to 
advertise a transmission to node $n_6$ in the second contention round, it would 
transmit bits 00 on SC $s_1$ (to advertise itself) and bits 01 on SC $s_4$ (to 
advertise the intended receiver).

\begin{table}[!t]
\caption{Example of extended SC mapping}
\label{tab:example_sc}
\centering
\begin{tabular}{l|c|c|c|c}
	\toprule
	\bfseries Node & \multicolumn{2}{c|}{$\mathcal{S}_1$} 
	& \multicolumn{2}{c}{$\mathcal{S}_2$} \\
	\cmidrule{2-3} \cmidrule{4-5}
	& SC Number & 
	Data on SC & SC Number & Data on SC \\
	\midrule
	$n_1$ & $s_1$ & 00 & $s_3$ & 00\\
	$n_2$ & $s_1$ & 01 & $s_3$ & 01\\
	$n_3$ & $s_1$ & 10 & $s_3$ & 10\\
	$n_4$ & $s_1$ & 11 & $s_3$ & 11\\				
	$n_5$ & $s_2$ & 00 & $s_4$ & 00\\
	$n_6$ & $s_2$ & 01 & $s_4$ & 01\\
	$n_7$ & $s_2$ & 10 & $s_4$ & 10\\
	$n_8$ & $s_2$ & 11 & $s_4$ & 11\\				
	\bottomrule
\end{tabular}
\end{table}

\rev{Another possible issue of the proposed subcarrier mapping is that it must 
be established at network setup, representing a problem in dynamic ad hoc 
networks where nodes join and leave continuously. To overcome this issue, each 
node should 
keep track of the first available slots in the maps $\Fa$ and $\Fb$. 
Whenever a node leaves the network, it should send a broadcast message 
indicating its slots, so that all remaining nodes mark them as free and update 
the information on the first available slots. When a node joins the network, 
conversely, it sends a broadcast message and waits for a reply, which will 
assign it the first available slots. In networks with multiple collision 
domains, the broadcast messages need to be propagated so that all nodes update 
the information and share the same version of the maps. Such a strategy will 
work with minor overhead if the network is not too dynamic. }

\subsection{Asynchronous channel access}
\label{subsec:asynchronous}

An important assumption that was made in Section~\ref{subsec:assumptions} is 
that the channel access is synchronous, i.e., all nodes try to access the 
channel at the same time. This is not realistic, since in real networks nodes 
often generate packets, and therefore try to access the channel, in an 
independent manner. 
As a consequence, when the proposed algorithm is implemented in a network 
with multiple collision domains, a node may start a 
contention procedure while another node within its range is receiving 
data, thus causing a collision. Indeed, the scanning procedure performed before 
the contention rounds is only capable of determining if a surrounding node is 
transmitting, not if it is receiving. 

\begin{figure}[!t]
	\centering
	\subfloat[Standard procedure: collision]{
	\scalebox{0.7}{
	\begin{tikzpicture}
	\draw[fill=green!70!white,draw=black] (2,-1) circle (8pt);	
	\node at (2,-1) {$n_3$};
	\draw[dashed,draw=green!70!white] (2,-1) circle (50pt);
	\draw[fill=red!50!white,draw=black] (1,0) circle (8pt);
	\node at (1,0) {$n_2$};
	\draw[fill=blue!50!white,draw=black] (0,1) circle (8pt);
	\node at (0,1) {$n_1$};
	\draw[dashed,draw=blue!50!white] (0,1) circle (50pt);
	\draw[-latex,draw=blue!50!white,fill=blue!50!white,very thick] (.2,.8) -- 
	(.8,.2);
	\draw[-latex,draw=green!70!white,fill=green!70!white,very thick,dashed] 
	(1.8,-.8) -- 
	(1.2,-.2);
	\node at (.75,-.6) { Collision};
	\end{tikzpicture}}
	\label{fig:example_collision}}
	\subfloat[Deferring: no collision]{
	\scalebox{0.72}{
	\begin{tikzpicture}
	\draw[fill=green!70!white,draw=black] (2,-1) circle (8pt);	
	\node at (2,-1) {$n_3$};
	\draw[dashed,draw=green!70!white] (2,-1) circle (50pt);
	\draw[fill=red!50!white,draw=black] (1,0) circle (8pt);
	\node at (1,0) {$n_2$};
	\draw[fill=blue!50!white,draw=black] (0,1) circle (8pt);
	\node at (0,1) {$n_1$};
	\draw[dashed,draw=blue!50!white] (0,1) circle (50pt);
	\draw[-latex,draw=blue!50!white,fill=blue!50!white,very thick] (.15,.75) -- 
	(.75,.15);
	\draw[latex-,draw=red!50!white,fill=red!50!white,very thick,dashed] 
	(1.85,-.75) -- (1.25,-.15);
	\draw[latex-,draw=red!50!white,fill=red!50!white,very thick,dashed] 
	(.25,.85) -- (.85,.25);
	\node at (3.2,-1) {Deferring!};
	\node at (.6,-.6) {No collision};
	\node at (1,.7) { \textcolor{red}{CTS}};
	\node at (1.9,-.25) { \textcolor{red}{CTS}};
	\end{tikzpicture}}
	\label{fig:example_deferring}}
	\caption{Example scenario of asynchronous channel access with potential	
	collisions.}
\end{figure}
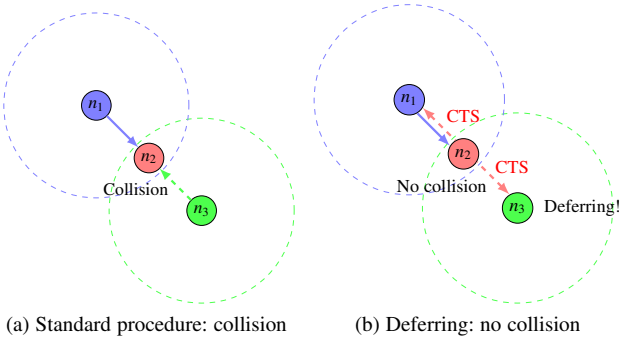

Fig.~\ref{fig:example_collision} reports an example of such a situation, where 
node $n_3$ tries to access the channel while node $n_1$ is already performing a 
data transmission to $n_2$, which is inside the coverage range of both nodes. 
When $n_3$ starts the first transmission round, it causes a collision with the 
ongoing transmission.

To cope with this issue, we make a simple yet effective modification to the 
algorithm presented in Section~\ref{sec:protocol}, so that an idle node (i.e., 
a node that does not have a packet to send, such as $n_3$ in 
Fig.~\ref{fig:example_deferring}) which hears a CTS from a 
neighboring node refrains from accessing the channel until the end of the 
transmission is advertised through an ACK packet. 
To prevent freezing (in case the ACK is lost), a timeout can be started upon 
CTS detection and the node can again access the channel after its expiration. 
Fig.~\ref{fig:example_deferring} shows that, if such a deferring policy is 
adopted, no collision happens in the previously described scenario.

\subsection{Impact of fading\rev{, lock problems and collisions}}
\label{subsec:fading}

In all the discussions so far we have assumed an ideal channel. Real wireless 
communication environments are characterized by impairments such as fading, 
shadowing and path loss. For our scheme, the case of selective fading, 
in which only narrow portions of the spectrum (corresponding to one or few 
subcarriers) are disturbed, is particularly challenging. Such a phenomenon 
could lead to sub--channel outage and the emergence of \textit{false negatives} 
(FNs), i.e., missed detection of data on a subcarrier \cite{Sen2011}. 

The impact of FNs in the three contention rounds of RCFD  can be summarized as 
follows:
\begin{enumerate}
\item \textit{First round:} Multiple PTs can be selected in the same collision 
domain as a result of FNs; as a consequence, nodes that should be RR in the 
second round would be PT instead and would not send the CTS in the third round, 
thus leading to missed transmission opportunities.
\item \textit{Second round:} A FN during the second round could lead to a node 
not receiving an RTS destined to it, again resulting in a missed opportunity 
for a transmission which, however, should have been authorized.
\item \textit{Third round:} Again, a FN occurrence during the third round 
results in a missed CTS reception and a corresponding missed transmission 
opportunity.
\end{enumerate}

In conclusion, FNs induced by sub--channel outage never result in a collision 
but only in possible missed transmission opportunities, thus causing 
underutilization of the channel and slightly degrading the efficiency of the 
protocol.

\rev{Similarly, channel underutilization may be caused by ``lock'' problems 
that arise for particular selections of subcarriers in the first 
round.\footnote{\rev{For example, consider a ``line'' network where adjacent 
nodes are in the same collision domain and they select SCs in the first round 
in ascending order. Only the first node will be the PT and transmit, while 
other concurrent transmissions may have been allowed.}} However, a different 
random selection is carried out at each transmission opportunity, thus 
preventing permanent lock problems. In general, RCFD is designed to ensure that 
collisions are avoided, at the cost of losing a transmission opportunity every 
now and then. 
The simulations of Section~\ref{sec:simulation_ns3} will demonstrate the 
effectiveness of this choice.}

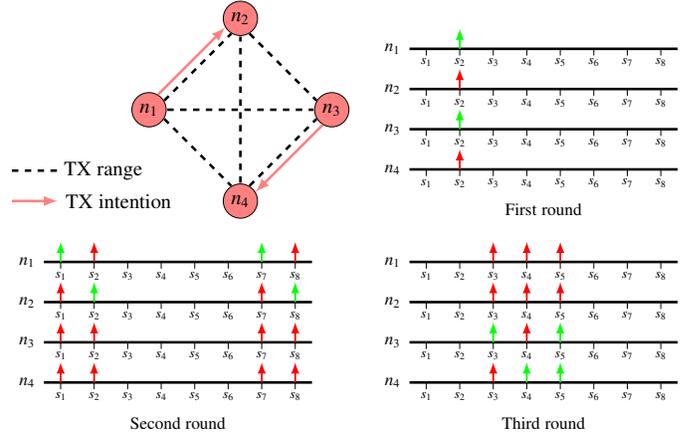
\begin{figure}[t!]
\begin{center}
\scalebox{0.81}{
\begin{tikzpicture}
	\begin{scope} 
		\draw[fill=red!50!white,draw=black] (0,-1) circle (8pt);
		\draw[fill=red!50!white,draw=black] (3,-1) circle (8pt);
		\draw[fill=red!50!white,draw=black] (1.5,0.5) circle (8pt);
		\draw[fill=red!50!white,draw=black] (1.5,-2.5) circle (8pt);
		\draw[dashed,very thick] (0.25,-.85) -- (1.35,0.25);
		\draw[dashed,very thick] (1.65,0.25) -- (2.75,-.85);
	    \draw[dashed,very thick] (0.25,-1.15) -- (1.35,-2.25);
	    \draw[dashed,very thick] (1.65,-2.25) -- (2.75,-1.15);	
	    \draw[dashed,very thick] (1.5,-2.2) -- (1.5,.2);	
	    \draw[dashed,very thick] (.3,-1) -- (2.7,-1);	  	
		\draw[-latex,draw=red!50!white,fill=red!50!white,very thick] 
		(0.15,-.75) -- (1.25,0.35);
		\draw[-latex,draw=red!50!white,fill=red!50!white,very thick] 
		(2.85,-1.25) -- (1.75,-2.35);
		\node at (0,-1) {$n_1$};
		\node at (1.5,0.5) {$n_2$};
		\node at (3,-1) {$n_3$};
		\node at (1.5,-2.5) {$n_4$};
		\begin{scope}[shift={(-5.5,-2.5)}]
			\draw [dashed, very thick] (3.25,0.5) -- (4,0.5);
			\draw [-latex, draw=red!50!white,fill=red!50!white,very thick] 
			(3.25,0) -- (4,0);
			\node at (4.75,0.5) [align=left] {TX range};
			\node at (5,0) [align=left] {TX intention};
		\end{scope}
	\end{scope}
	\begin{scope}[shift={(4,0)},scale=0.55] 
		\foreach[count=\i] \y in {0,-1.2,-2.4,-3.6}{
			\node[font=\footnotesize] at (0,\y) {$n_{\i}$};
			\draw[draw=black,very thick] (.5,\y)--(8.5,\y);
			\foreach \x in {1,2,...,8}{
				\draw (\x,\y) -- (\x,\y-.2);
				\node[font=\footnotesize] at (\x,\y-.4) 
				{\scalebox{0.8}{$s_{\x}$}};
			}
		}
		\draw[-latex, thick,draw=green,fill=green] (2,0) -- (2,.6);
		\draw[-latex, thick,draw=green,fill=green] (2,-2.4) -- (2,-1.8);
		\draw[-latex, thick,draw=red,fill=red] (2,-1.2) -- (2,-.6);
		\draw[-latex, thick,draw=red,fill=red] (2,-3.6) -- (2,-3);
		\node[font=\footnotesize] at (4.5,-4.8) {First round};
	\end{scope}
	\begin{scope}[shift={(-2,-3.5)},scale=0.55] 
		\foreach[count=\i] \y in {0,-1.2,-2.4,-3.6}{
			\node[font=\footnotesize] at (0,\y) {$n_{\i}$};
			\draw[draw=black,very thick] (.5,\y)--(8.5,\y);
			\foreach \x in {1,2,...,8}{
				\draw (\x,\y) -- (\x,\y-.2);
				\node[font=\footnotesize] at (\x,\y-.4) 
				{\scalebox{0.8}{$s_{\x}$}};
			}
		}
		\draw[-latex, thick,draw=green,fill=green] (1,0) -- (1,.6);
		\draw[-latex, thick,draw=green,fill=green] (7,0) -- (7,.6);
		\draw[-latex, thick,draw=red,fill=red] (2,0) -- (2,.6);
		\draw[-latex, thick,draw=red,fill=red] (8,0) -- (8,.6);
		\draw[-latex, thick,draw=green,fill=green] (2,-1.2) -- (2,-.6);
		\draw[-latex, thick,draw=green,fill=green] (8,-1.2) -- (8,-.6);
		\draw[-latex, thick,draw=red,fill=red] (1,-1.2) -- (1,-.6);
		\draw[-latex, thick,draw=red,fill=red] (7,-1.2) -- (7,-.6);
		\draw[-latex, thick,draw=red,fill=red] (2,-2.4) -- (2,-1.8);
		\draw[-latex, thick,draw=red,fill=red] (8,-2.4) -- (8,-1.8);
		\draw[-latex, thick,draw=red,fill=red] (1,-2.4) -- (1,-1.8);
		\draw[-latex, thick,draw=red,fill=red] (7,-2.4) -- (7,-1.8);
		\draw[-latex, thick,draw=red,fill=red] (2,-3.6) -- (2,-3);
		\draw[-latex, thick,draw=red,fill=red] (8,-3.6) -- (8,-3);
		\draw[-latex, thick,draw=red,fill=red] (1,-3.6) -- (1,-3);
		\draw[-latex, thick,draw=red,fill=red] (7,-3.6) -- (7,-3);
		\node[font=\footnotesize] at (4.5,-4.8) {Second round};				
	\end{scope}
	\begin{scope}[shift={(4,-3.5)},scale=0.55] 
		\foreach[count=\i] \y in {0,-1.2,-2.4,-3.6}{
			\node[font=\footnotesize] at (0,\y) {$n_{\i}$};
			\draw[draw=black,very thick] (.5,\y)--(8.5,\y);
			\foreach \x in {1,2,...,8}{
				\draw (\x,\y) -- (\x,\y-.2);
				\node[font=\footnotesize] at (\x,\y-.4) 
				{\scalebox{0.8}{$s_{\x}$}};
			}
		}
		\draw[-latex, thick,draw=red,fill=red] (3,0) -- (3,.6);
		\draw[-latex, thick,draw=red,fill=red] (4,0) -- (4,.6);
		\draw[-latex, thick,draw=red,fill=red] (5,0) -- (5,.6);
		\draw[-latex, thick,draw=red,fill=red] (3,-1.2) -- (3,-.6);
		\draw[-latex, thick,draw=red,fill=red] (4,-1.2) -- (4,-.6);
		\draw[-latex, thick,draw=red,fill=red] (5,-1.2) -- (5,-.6);
		\draw[-latex, thick,draw=green,fill=green] (3,-2.4) -- (3,-1.8);
		\draw[-latex, thick,draw=green,fill=green] (5,-2.4) -- (5,-1.8);
		\draw[-latex, thick,draw=red,fill=red] (4,-2.4) -- (4,-1.8);
		\draw[-latex, thick,draw=green,fill=green] (4,-3.6) -- (4,-3);
		\draw[-latex, thick,draw=green,fill=green] (5,-3.6) -- (5,-3);
		\draw[-latex, thick,draw=red,fill=red] (3,-3.6) -- (3,-3);
		\node[font=\footnotesize] at (4.5,-4.8) {Third round};				
	\end{scope}
\end{tikzpicture}}
\end{center}
\caption{\rev{Example of a 4--node network in one collision domain where two 
nodes ($n_1$ and $n_3$) become PT simultaneously because they select the same 
subcarrier in the first round ($s_2$). The contention is solved in the third 
round where only $n_1$ receives the CTS and, hence, is cleared to transmit.}}
\label{fig:first_round_collision}
\end{figure}

\rev{Another possible issue arises in RCFD when multiple nodes select 
the same SC in the first contention round, when the SC choice is random. This 
represents a problem in the BACK2F scheme \cite{Sen2011}, that was addressed by 
performing multiple rounds but still maintaining a residual collision 
probability, as reported in Section~\ref{sec:theoretical}. Conversely, in our 
protocol, this could result in multiple PTs being present, only one of which is 
selected in the following rounds, thus preventing any possible collision. 
An example of how RCFD handles the issue of multiple PTs in the same 
collision domain is provided in Fig.~\ref{fig:first_round_collision}.}

\rev{Finally, real--world implementations of FD devices likely do not achieve 
perfect SI cancellation and may be impaired by residual self--interference. If 
this interference is too high, it can impact RCFD in two ways. First, every 
bidirectional FD transmission will be less robust, leading to lower overall 
performance. Second, the detection of SCs in the three contention rounds for a 
node that is also transmitting in one or more rounds may be more difficult, 
and false negatives such as those described at the beginning of this subsection 
may occur. However, it has to be noted that working implementations of FD 
devices able to reduce the residual SI to the noise floor can be found in the 
literature \cite{Bharadia2013}. Nevertheless, future activities are planned to 
assess the performance of RCFD under different levels of residual SI.}

\subsection{Possible protocol improvements}
\label{subsec:optimization}

The RCFD protocol in the presented form already yields significant performance 
benefits, as will be shown in Sections~\ref{sec:theoretical} and 
\ref{sec:simulation_ns3}.

Further improvements in channel utilization can be achieved if the ACK 
procedure is also moved to the frequency domain, as already suggested in 
\cite{Zhang2012}. 
The implementation of this enhancement would be straightforward, since a 
mapping between nodes and subcarriers is already established.

Moreover, as discussed in \cite{Sen2010} and \cite{Sen2011}, the random 
selection of  OFDM subcarriers implicitly defines an order among the nodes
trying to access the channel, thus enabling the possibility of fast and 
efficient TDMA--like transmissions. 
Alternatively, unlike we assumed throughout this paper, the order among 
nodes can be exploited if the nodes in the network have different priorities. 
In this case, the first round of the RCFD algorithm can be modified by letting 
a high priority node randomly choose its SC among a subset of $\SC$ which 
contains lower frequency SCs with respect to the set in which a low priority 
node picks its SC. This would guarantee to the former node a higher 
probability of being selected as a PT and, hence, a faster channel access.

\rev{For the sake of clarity, the version of RCFD evaluated in the next 
sections does not include these improvements, whose detailed design and 
performance evaluation are left for future research. However, it does include 
the extended SC mapping described in 
Section~\ref{subsec:mapping}, as well as the deferring policy to allow 
asynchronous access scheme described in Section~\ref{subsec:asynchronous}.}

\section{Theoretical analysis}
\label{sec:theoretical}

In order to validate the proposed protocol and highlight the benefits it is 
able to provide, we compare its performance against those offered by 
standard MAC algorithms for wireless networks and other state--of--the--art 
strategies.

In this section we provide a theoretical comparison based on the analytical 
evaluation of the \textit{normalized saturation throughput} of different MAC 
algorithms. This quantity is defined as the maximum load that a system is able 
to carry without becoming unstable \cite{Bianchi2000}. It can also be seen as 
the percentage of time in which nodes with full buffers utilize the 
channel for data transmission using a contention--based MAC scheme. In order to 
make the problem analytically tractable, some assumptions are made. We consider 
a network of $N$ nodes, all within the 
same collision domain and with saturated queues, meaning that every node always 
has at least one packet to transmit. A First In First Out (FIFO) policy is 
adopted at each node, meaning that only the packet at the head of the 
queue can be transmitted. Furthermore, an ideal communication channel is 
assumed, so that the only cause of transmission errors would be collisions 
among different packets. \rev{In the case of frequency--based channel access 
schemes, we assumed that the exchange of data on subcarriers during the 
contention round works perfectly, regardless of the number of nodes in the 
network (if $N>S/2$ we can assume that the extended mapping scheme of 
Section~\ref{subsec:mapping} is adopted).} Finally, we suppose that both the 
transmission rate $R$
and the payload size $L$ (in Bytes) are fixed.

We consider four different MAC layer protocols to compare with our proposed 
RCFD strategy. The baseline scheme is the IEEE 802.11 Distributed Coordination 
Function (DCF) proposed in the standard \cite{ieee80211std}, both with and 
without the RTS/CTS option.  We selected the FD MAC strategy \cite{Duarte2014} 
among the various time--domain MAC protocols for FD networks discussed in 
Section~\ref{subsec:related_fd}, since it is one of the most general 
approaches, and does not impose any assumption on network topology, traffic 
pattern or PHY configuration. Finally, the BACK2F scheme \cite{Sen2011} has 
been chosen as a protocol that performs channel contention in the frequency 
domain. 

In order to obtain a fair comparison, all the protocols are based on the same 
underlying physical layer, specifically that described by the IEEE 802.11g 
standard, which is very widespread. 
Tab.~\ref{tab:analysis_param} reports the main parameters considered in this 
theoretical analysis.

\begin{table}[!b]
\caption{System parameters for theoretical analysis}
\label{tab:analysis_param}
\centering
\scalebox{0.9}{
\begin{tabular}{l|c|c}
\toprule
{\bfseries Parameter} & {\bfseries Description} & {\bfseries Value}\\
\midrule
$T_{ack}$ & MAC--layer ACK transmission time & 50\mus\\
$T_{rts}$ & RTS frame transmission time & 58\mus\\
$T_{cts}$ & CTS frame transmission time & 50\mus\\
$T_{sifs}$ & Short Inter--Frame Space & 10\mus\\
$T_{difs}$ & DCF Inter--Frame Space & 28\mus\\
$T_{p}$ & Propagation time over the air & 1\mus\\
$T_{slot}$ & MAC--layer slot time & 9\mus\\
$W$ & Initial value of backoff window & 16\\
$m$ & Maximum number of retransmission attempts & 6\\
$S$ & Number of available OFDM subcarriers & 52\\
$T_{round}$ & Duration of a contention round in the frequency domain & 6\mus\\
\bottomrule
\end{tabular}}
\end{table}

\subsection{Analysis for IEEE 802.11 and FD MAC}

The starting point for the analysis is the work in \cite{Bianchi2000}, where 
the normalized saturation throughput was derived for the IEEE 802.11 DCF (with 
and without RTS/CTS). In this section we report the main results of that study 
and extend them to evaluate the normalized saturation throughput for the FD MAC 
algorithm \cite{Duarte2014}. 

We recall that the IEEE 802.11 DCF is based on a CSMA/CA strategy, where nodes 
listen to the channel before transmitting. If they find it busy, they wait 
until it becomes idle, and then defer transmission for an additional random backoff period in order to avoid collisions.
The first analysis step is, hence, the introduction of a discrete--time Markov 
model to describe the behavior of a single station during backoff periods. 
This model was then used to derive the probability $\tau$ that a single station 
transmits in a randomly chosen slot and the probability $p$ that a transmission 
results in a collision, as functions of the system parameters, such as the 
initial value of the backoff window $W$ and the maximum number of backoff 
stages $m$. 
Subsequently, two probabilities were computed, namely $P_{tr}$, the probability 
that at least a transmission attempt takes place in a slot, and $P_s$, the 
probability that this transmission is successful, expressed as functions of the 
number of nodes in the network $N$, and of the probabilities $\tau$ and $p$.
Specifically, the number of stations that transmit in a given slot is a 
binomial random variable $B$ of parameters $N$ and $\tau$ and the 
probabilities $P_{tr}$ and $P_s$ can be expressed as
\begin{align}
P_{tr} &= P\left(B\geq1\right)=1-\left(1-\tau\right)^N\label{eq:Ptr}\\
P_s &= P\left(B=1\vert B\geq 
1\right)=\frac{N\tau\left(1-\tau\right)^{N-1}}{1-\left(1-\tau\right)^N}\label{eq:Ps}
\end{align}

Finally, the saturation throughput can be computed as
\begin{equation}
\label{eq:eta_wifi}
\eta_{DCF}=\frac{P_{tr}P_sT_d}{\left(1-P_{tr}\right)T_{slot}+P_{tr}P_sT_S+P_{tr}\left(1-P_s\right)T_C}
\end{equation}
where $T_d$ is the payload transmission time, $T_{slot}$ is the slot time in 
IEEE 802.11, $T_S$ is the slot duration in case of a successful transmission 
and $T_C$ is the slot duration in case of a collision. The values for $T_S$ and 
$T_C$, as computed in \cite{Bianchi2000}, are 
\begin{align}
T_S &= T_{difs}+T_d+T_{sifs}+T_{ack}+2T_p\label{eq:TS}\\
T_C &= T_{difs}+T_d+T_p\label{eq:TC}
\end{align}
for the standard IEEE 802.11 DCF without RTS/CTS and
\begin{align}
T_S &= 
T_{difs}+T_{rts}+T_{cts}+T_d+3T_{sifs}+T_{ack}+4T_p\label{eq:TSrts}\\
T_C &= T_{difs}+T_{rts}+T_p\label{eq:TCrts}
\end{align}
in case the RTS/CTS option is enabled. The meaning and the values of parameters 
$T_{difs}$, $T_{sifs}$, $T_{rts}$, $T_{cts}$, $T_{ack}$ and $T_p$ are 
reported in Tab.~\ref{tab:analysis_param}, whereas the transmission 
time $T_d$ for a packet of length $L$ sent at rate $R$ can be derived from 
the IEEE 802.11 specifications \cite{ieee80211std}.

\rev{The FD MAC algorithm, presented in \cite{Duarte2014}, builds on the IEEE 
802.11 DCF with the use of RTS and CTS frames, with a substantial difference: 
when node $n_j$ receives an RTS from node $n_i$, it checks at the head of its 
transmission queue if there is a packet destined to $n_i$ and, if present, 
starts to transmit it immediately after the CTS frame, with a waiting period of 
$T_{sifs}$. Other minor modifications to the DCF include the possibility for a 
node to receive both a data frame and an ACK frame within a network allocation 
vector (NAV) interval and the possibility to send an ACK while waiting for 
another ACK \cite{Duarte2014}.} 

The analysis presented for the IEEE 802.11 DCF in \cite{Bianchi2000} is 
extended to account also for the FD MAC, taking into account that a 
successful FD transmission can occur in two different cases.
The first one is when only two nodes grab the channel simultaneously and have 
packets for each other, which happens with probability 
\begin{equation}
\frac{P\left(B=2\vert B\geq 1\right)}{\left(N-1\right)^2}=
\frac{N\tau^2\left(1-\tau\right)^{N-2}}{2\left(N-1\right)\left(1-\left(1-\tau\right)^N\right)}
\end{equation}
since the probability that a generic node has a packet for another specific 
node is $1/\left(N-1\right)$.
A successful FD communication takes place also if a single node grabs the 
channel, which happens with probability expressed by \eqref{eq:Ps}, and the 
target receiver has a packet for it at the head of the queue, which happens with probability 
$1/\left(N-1\right)$.
Hence, the probability that a successful FD transmission takes place is given by 
\begin{equation}
\small
P_{s,fd}=\frac{P\left(B=2\vert B\geq 
1\right)}{\left(N-1\right)^2}+\frac{P\left(B=1\vert B\geq 
1\right)}{N-1}=\frac{N\tau\left(1-\tau\right)^{N-2}\left(2-\tau\right)}{2\left(N-1\right)\left(1-\left(1-\tau\right)^N\right)}
\end{equation}
A successful HD transmission happens when a single node grabs the 
channel but the target receiver does not have a packet for it, which occurs 
with probability
\begin{equation}
\small
P_{s,hd}=P\left(B=1\vert B\geq 
1\right)\left(1-\frac{1}{N-1}\right)=
\frac{N\left(N-2\right)\tau\left(1-\tau\right)^{N-1}}{\left(N-1\right)\left(1-\left(1-\tau\right)^N\right)}
\end{equation}
Consequently, the saturation throughput is given by
\begin{equation}
\label{eq:eta_fd}
\eta_{FD} = 
\frac{T_dP_{tr}P_{s,hd}+2T_dP_{tr}P_{s,fd}}{\left(1-P_{tr}\right)T_{slot}+P_{tr}P_sT_S+P_{tr}\left(1-P_s\right)T_C}
\end{equation}
where $P_{tr}$, $T_S$ and $T_C$ are expressed by 
\eqref{eq:Ptr},~(\ref{eq:TSrts}) and (\ref{eq:TCrts}) respectively.

{\color{black}
\subsection{Analysis for BACK2F}
\label{subsec:back2f_analysis}

The Markov model introduced in \cite{Bianchi2000} is no longer useful with 
the BACK2F scheme described in \cite{Sen2011}. 
In this channel access scheme, indeed, there cannot be any idle slots (i.e., 
$P_{tr}=1$) and the only case in which a transmission is not successful is when 
there is a collision on the SC selection after the second contention round in 
the frequency domain.
An original Markov model is introduced in this section to derive the success 
probability $P_S$, i.e., the probability that no collisions happen, as a 
function of the number of nodes $N$ and the number of available OFDM 
subcarriers $S$. 

Specifically, we consider a discrete--time Markov chain that models the 
three-dimensional process $\left\{x(t),c(t),y(t)\right\}$, where $x(t)$ 
represents the number 
of nodes winning the first contention round of BACK2F in time slot $t$, $c(t)$ 
represents the lowest--frequency SC during the first contention round in the 
same time slot and $y(t)$ represents the number of nodes winning the second 
contention round. The processes $x(t)$ and $y(t)$ take values in the set 
$\left\{1,\dots,N\right\}$, while $c(t)$ can range from $0$ to 
$S-1$.\footnote{\rev{We assume without loss of generality that 
$\mathcal{S}=\left\{0,1,\dots,S-1\right\}$.}} Trivially, it must 
hold $y(t)\leq x(t)$, since only the nodes that have won the first round can 
take part in the second one, and also $x(t)=N$ if $c(t)=S-1$. Moreover, if 
$c(t)=S-1$, it means that all the nodes have won the first contention round, 
i.e., $x(t)=N$. Taking these constraints into account, the number of reachable 
states is $N\cdot\left(N-1\right)\cdot\left(S-1\right)/2+N$.

It can be proved that the proposed chain is time--homogeneous, irreducible and 
aperiodic and, hence, a stationary distribution can be found as 
\begin{equation}
\pi_{i,a,j}=\lim\limits_{t\to\infty}P\left\{x(t)=i,c(t)=a,y(t)=j\right\}
\end{equation}
for $i=1,\dots,N$, $a=0,\dots,S-1$ and $j=1,\dots,i$. The stationary 
distribution is derived from the transition probabilities between the different 
states, which are computed in detail in Appendix~\ref{sec:appendix}.

A collision in a time slot can happen only if two or more nodes win the second 
contention round, i.e., if $y(t)>1$. The success probability can hence be 
computed as
\begin{equation}
P_s=\sum\limits_{i=1}^N\sum\limits_{a=0}^{S-1}\pi_{i,a,1}
\end{equation}

Once this probability is obtained, the saturation throughput is given by
\begin{equation}
\label{eq:eta_b2f}
\eta_{B2F} = 
\frac{P_sT_d}{P_sT_S+\left(1-P_s\right)T_C}
\end{equation}
Considering the structure of the BACK2F protocol, the values of $T_S$ and 
$T_C$ are equal to
\begin{align}
T_S &= T_{difs} + 2T_{round} + T_d + T_{sifs} + T_{ack} + 2T_p\\
T_C &= T_{difs} + 2T_{round} + T_d + T_p
\end{align}
where $T_{round}$ is the duration of a contention round in the frequency 
domain, reported in Tab.~\ref{tab:analysis_param}.
}

\subsection{\rev{Analysis for RCFD}}

In the RCFD protocol, similarly to what happens in BACK2F, there are no 
idle slots. Moreover, the RTS/CTS exchange in the frequency domain prevents any 
possibility of collision.
As a consequence, we have $P_{tr}=1$ and $P_s=P_{s,hd}+P_{s,fd}=1$, where
\begin{equation}
P_{s,hd} = 1-\frac{1}{N-1},\quad P_{s,fd}=\frac{1}{N-1}
\end{equation}
The saturation throughput hence becomes
\begin{equation}
\label{eq:eta_rcfd}
\eta_{RCFD} = 
\frac{T_dP_{s,hd}+2T_dP_{s,fd}}{T_S}
\end{equation}
where, in this case
\begin{equation}
T_S = T_{difs} + 3T_{round} + T_h + T_d + T_{sifs} + T_{ack} + 2T_p
\end{equation}
In both this analysis and the one of BACK2F we have considered the scanning 
time $T_{scan}$ used in \eqref{eq:channel_access} equal to $T_{difs}$, to 
provide a fair comparison among all the MAC protocols.

\subsection{Numerical results}

\begin{figure}[!t]
    \centering
    \includegraphics[width=0.9\columnwidth]{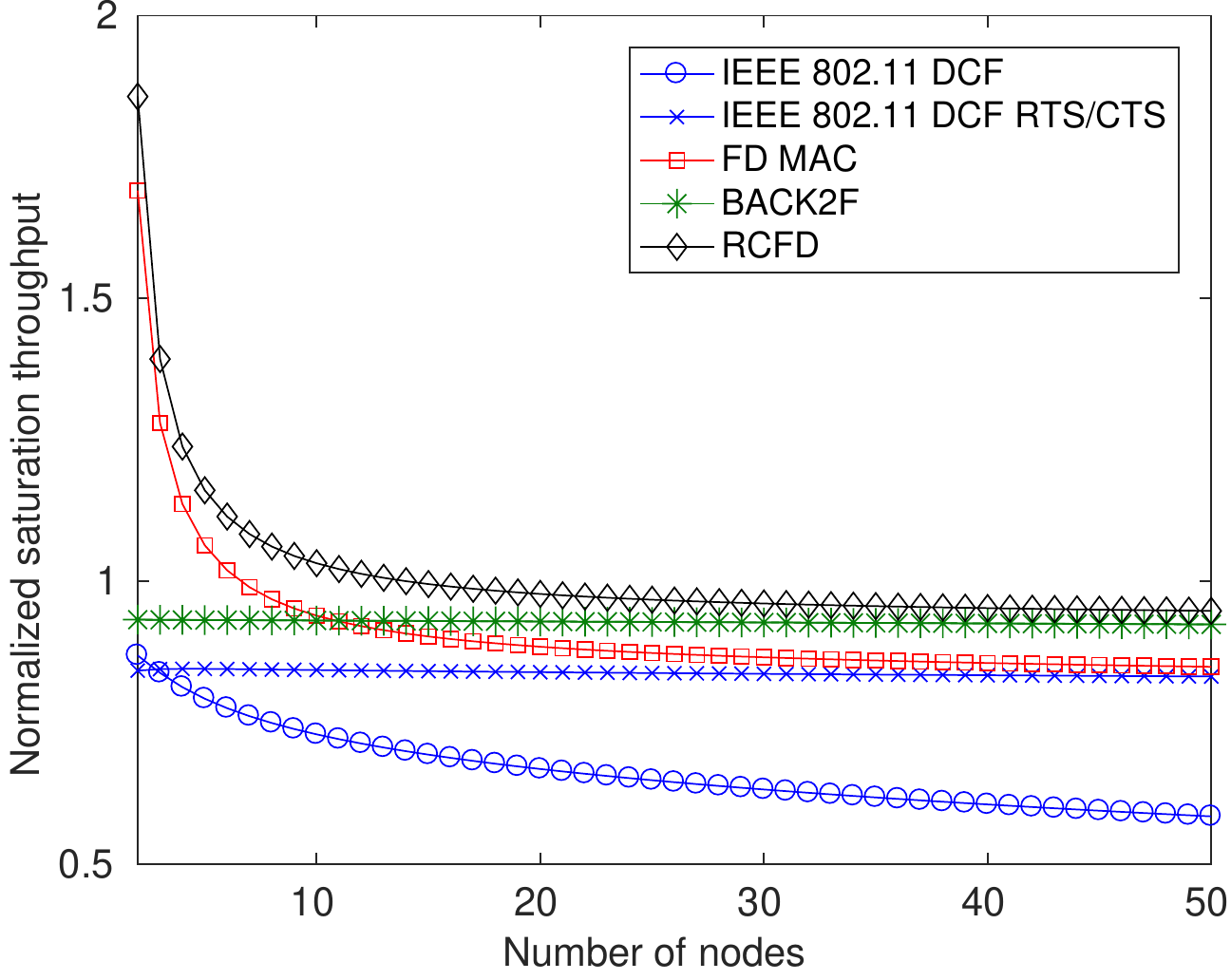}
    \caption{Theoretical saturation throughput versus number of nodes, 
    with L=1000~Bytes long packets and R=6~Mbps data rate.}
    \label{fig:analytic_result_nodes}
\end{figure}  

In the previous subsection we have derived the saturation throughput for the 
different MAC protocols as a function of several system parameters. We will 
now numerically evaluate this metric for different network configurations and 
system parameters. 
Tab.~\ref{tab:analysis_param} reports the simulation parameters in this 
evaluation, which are adopted from the IEEE 802.11g standard 
\cite{ieee80211std}.

Fig.~\ref{fig:analytic_result_nodes} shows the saturation throughput for all 
MAC algorithms versus the number of nodes in the network. 
The payload length has been kept fixed at $L=1000$~Bytes, while the 
transmission rate is $R=6$~Mbps, yielding a data transmission time of roughly 
$T_d=1.4$~ms. 
It can be observed that the RCFD strategy outperforms all other MAC algorithms 
for any number of nodes. The two schemes that consider FD transmissions (RCFD 
and FD MAC) are able to provide a normalized throughput higher than one, for a 
small number of nodes. BACK2F and IEEE 802.11 RTS/CTS do not show a significant 
variation with the number of nodes, with the first one providing a higher 
throughput (close to 1) and performing close to RCFD for a large number of 
nodes. The standard IEEE 802.11 DCF provides the worst performance, 
strongly affected by the number of nodes, as expected.

\rev{It is worth noting that the sharp decrease in throughput presented by 
FD--capable MAC protocols (RCFD and FD MAC) is due to the FIFO assumption. 
Indeed, in both cases, assuming that a node $n_i$ gets the channel, a FD 
transmission happens only if the packet at the head of the queue of the 
receiver $n_j$ is destined to $n_i$, which happens with probability $1/(N-1)$. 
The throughput curves for these algorithm, hence, follow a hyperbolic shape.
The FIFO assumption was considered in this analysis for the sake of 
tractability and will be relaxed in the simulations of 
Section~\ref{sec:simulation_ns3}.}

\begin{figure}[t!]
\centering
\includegraphics[width=0.9\columnwidth]{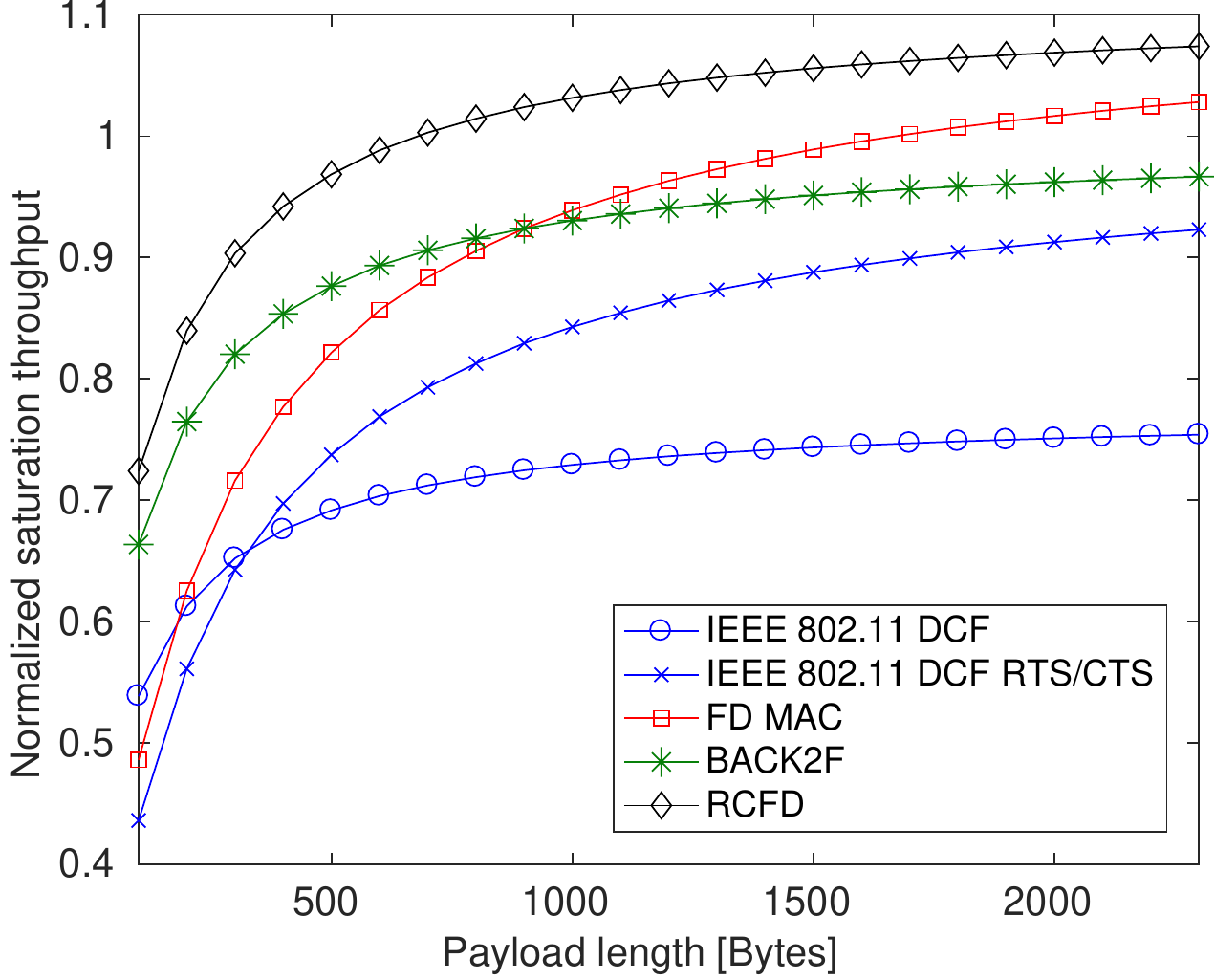}
\caption{Theoretical saturation throughput versus packet 
payload size for a network with $N=10$~nodes using an $R=6$~Mbps data 
rate.}
\label{fig:analytic_result_lengths}
\end{figure}

Another evaluation is reported in Fig.~\ref{fig:analytic_result_lengths}, where 
we kept the number of nodes and the data rate fixed at $N=10$ and 
$R=6$~Mbps, respectively, and we varied the payload length $L$
from 100 to 2300~Bytes. 
Again, the proposed RCFD technique provides the best performance for 
all possible payload sizes. The techniques based on time domain RTS/CTS (IEEE 
802.11 and FD MAC) perform very poorly for short packets, since in that 
case the overhead represented by the exchange of RTS and CTS frames has a very 
significant impact. The techniques that include frequency--based contention 
(RCFD and BACK2F) are characterized by a similar trend, even if the first one 
always provides a higher throughput, thanks to its FD capabilities. The 
standard IEEE 802.11 DCF without RTS/CTS, finally, yields the worst results, 
since it clearly suffers from the occurrence of collisions.

\rev{In order to make an assessment of the numerical results based on the 
theoretical models
presented in this section, a set of network simulations have been performed 
using the \textit{ns3} platform \cite{ns3},
configured according to the following assumptions:
\begin{itemize}
\item $N$ nodes are randomly deployed in the same collision domain;
\item each node randomly generates packets for every other node in the network 
and the transmission queue (which follows a FIFO behavior) is always saturated;
\item the communication channel is ideal, with collisions being the only source 
of errors;
\item the values of transmission rate ($R=6$~Mbps) and payload size 
($L=1000$~Bytes) are fixed.
\end{itemize}
The results, which refer to the simulation throughput averaged over 10 
different simulation runs, are reported in Tab.~\ref{tab:analysis_simulations}, 
where they are compared with the numerical values of 
Fig.~\ref{fig:analytic_result_nodes}.
It can be observed that the results of the analysis and simulations are 
close.  
Moreover, the simulations confirm that RCFD outperforms the other channel 
access schemes for any network size, as the analysis suggested.}

\begin{table}[!tb]
\caption{\rev{Comparison of normalized saturation throughput in analysis and 
simulations for FD, BACK2F and RCFD channel access schemes}}
\label{tab:analysis_simulations}
\centering
\scalebox{1}{
\begin{tabular}{>{\color{black}}l|>{\color{black}}c|>{\color{black}}c|>{\color{black}}c|>{\color{red}}c}
\toprule
\textbf{Algorithm} & $\mathbf{N=2}$ & $\mathbf{N=10}$ & $\mathbf{N=20}$ & 
$\mathbf{N=50}$\\
\midrule
FD Analysis & 1.6908 & 0.9390 & 0.8840 & 0.8485\\
FD Simulations & 1.4281 & 0.8458 & 0.7929 & 0.7377\\
\midrule
BACK2F Analysis & 0.9319 & 0.9304 & 0.9287 & 0.9235\\
BACK2F Simulations & 0.9312 & 0.8814 & 0.8280 & 0.7016\\
\midrule
RCFD Analysis & 1.8570 & 1.0316 & 0.9773 & 0.9474\\
RCFD Simulations & 1.8514 & 0.9306 & 0.9301 & 0.9300\\
\bottomrule
\end{tabular}}
\end{table}

\section{Simulation assessment}
\label{sec:simulation_ns3}

The results of Section~\ref{sec:theoretical} show a clear 
prevalence of the proposed RCFD algorithm over other MAC layer schemes considered. 
However, the analysis and simulations were conducted under some possibly 
limiting 
assumptions, the most important one being that all nodes are within the same 
collision domain.
In order to relax this assumption, the five aforementioned MAC strategies have 
been compared through \textit{ns3}, for the case of a wireless network 
with multiple collision domains.\footnote{MATLAB simulations were reported in 
the conference version of this paper \cite{Luvisotto2016}.} 

\subsection{Simulations setup}
\label{subsec:simulation_setup}

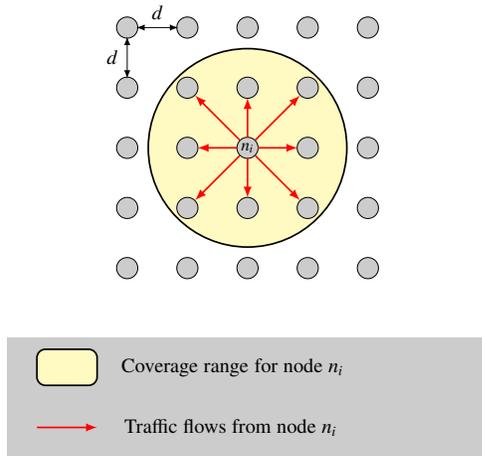
\begin{figure}[!bt]
\begin{center}
\scalebox{0.8}{
\begin{tikzpicture}
\draw[fill=yellow!30!white,rounded corners,thick] (2,2) circle (1.65);
\draw[fill=black!20!white,draw=black] (0,0) circle (5pt);
\draw[fill=black!20!white,draw=black] (1,0) circle (5pt);
\draw[fill=black!20!white,draw=black] (2,0) circle (5pt);
\draw[fill=black!20!white,draw=black] (3,0) circle (5pt);
\draw[fill=black!20!white,draw=black] (4,0) circle (5pt);
\draw[fill=black!20!white,draw=black] (0,1) circle (5pt);
\draw[fill=black!20!white,draw=black] (1,1) circle (5pt);
\draw[fill=black!20!white,draw=black] (2,1) circle (5pt);
\draw[fill=black!20!white,draw=black] (3,1) circle (5pt);
\draw[fill=black!20!white,draw=black] (4,1) circle (5pt);
\draw[fill=black!20!white,draw=black] (0,2) circle (5pt);
\draw[fill=black!20!white,draw=black] (1,2) circle (5pt);
\draw[fill=black!20!white,draw=black] (2,2) circle (5pt);
\draw[fill=black!20!white,draw=black] (3,2) circle (5pt);
\draw[fill=black!20!white,draw=black] (4,2) circle (5pt);
\draw[fill=black!20!white,draw=black] (0,3) circle (5pt);
\draw[fill=black!20!white,draw=black] (1,3) circle (5pt);
\draw[fill=black!20!white,draw=black] (2,3) circle (5pt);
\draw[fill=black!20!white,draw=black] (3,3) circle (5pt);
\draw[fill=black!20!white,draw=black] (4,3) circle (5pt);
\draw[fill=black!20!white,draw=black] (0,4) circle (5pt);
\draw[fill=black!20!white,draw=black] (1,4) circle (5pt);
\draw[fill=black!20!white,draw=black] (2,4) circle (5pt);
\draw[fill=black!20!white,draw=black] (3,4) circle (5pt);
\draw[fill=black!20!white,draw=black] (4,4) circle (5pt);
\draw[latex-latex] (.15,4) -- (.85,4);
\node at (.5,4.25) {$d$};
\draw[latex-latex] (0,3.85) -- (0,3.15);
\node at (-.25,3.5) {$d$};
\node at (2,2) {\footnotesize $n_i$};
\draw[-latex,draw=red, fill=red, thick] (2,2.17) -- (2,2.83);
\draw[-latex,draw=red, fill=red, thick] (2.17,2) -- (2.83,2);
\draw[-latex,draw=red, fill=red, thick] (2,1.83) -- (2,1.17);
\draw[-latex,draw=red, fill=red, thick] (1.83,2) -- (1.17,2);
\draw[-latex,draw=red, fill=red, thick] (2.13,2.13) -- (2.87,2.87);
\draw[-latex,draw=red, fill=red, thick] (2.13,1.87) -- (2.87,1.13);
\draw[-latex,draw=red, fill=red, thick] (1.87,1.87) -- (1.13,1.13);
\draw[-latex,draw=red, fill=red, thick] (1.87,2.13) -- (1.13,2.87);
\draw [draw=none,fill=black!20!white] (-2,-1.15) -- (6,-1.15) -- (6,-3.15) -- 
(-2,-3.15) -- (-2,-1.15);
\draw[fill=yellow!30!white,rounded corners,thick] (-1.5,-1.35) -- (-0.5,-1.35) 
-- (-0.5,-1.95) -- (-1.5,-1.95) -- cycle;
\node at (1.75,-1.65) {Coverage range for node $n_i$};
\draw[-latex,draw=red, fill=red,thick] (-1.5,-2.65) -- (-0.5,-2.65);
\node at (1.72,-2.65) {Traffic flows from node $n_i$};
\end{tikzpicture}}
\end{center}
\caption{\rev{Simulated network for the \textit{structured scenario}}.}
\label{fig:simulation_scenario}
\end{figure}

The standard distribution of \textit{ns3} already contains 
models for the IEEE 802.11 DCF, both with and without RTS/CTS, as defined in 
the standard. However, the modules for the MAC algorithms proposed in 
the literature, namely FD MAC, BACK2F and our proposal RCFD, were not 
available and therefore had to be purposely developed. Moreover, the standard \textit{ns3 wifi} 
module only allows half--duplex communications, preventing a node from 
transmitting if it is receiving. 
In order to be able to simulate a network with full--duplex nodes, we adopted 
the patch discussed in \cite{Zhou2014}, which allows to simulate an FD
wireless network with \textit{ns3}. It is worth stressing that, for the 
algorithms based on frequency domain operations (BACK2F and RCFD), the 
exchange of data over OFDM subcarriers during the contention rounds is assumed 
to be ideal, i.e., when a node transmits on a subcarrier all the other nodes in 
its collision domain are able to detect it.

\rev{Two different scenarios have been simulated: a \textit{structured 
scenario} and a \textit{random scenario}, described in detail in the following.}

\subsubsection{\rev{Setup for the structured scenario}}

The simulated network for the \textit{structured scenario} is depicted in 
Fig.~\ref{fig:simulation_scenario}. It is 
an ad hoc wireless network composed of fixed nodes placed on a grid. 
The distance between two adjacent nodes in the same row or column is $d$. The 
coverage range of each node is a circle of radius $r=d\sqrt{2}$ and, hence, 
includes all its one--hop neighbors. Within this area, the node can transmit 
and receive packets as well as overhear transmissions. 
To implement this channel model, the \textit{RangePropagationLossModel} of 
\textit{ns3} has been adopted\rev{, combined with a purposely implemented 
error model. According to these models, a transmission between two nodes is 
successful only if the distance is below $r$ and there is no collision, and it 
fails with probability 1 otherwise (regardless of the adopted transmission 
rate).}
\rev{In this way, the impact of collisions on the network can be accurately 
analyzed for the different channel access strategies, isolating it from all the 
other factors that can affect the performance, such as path loss, fading, 
performance of different modulation and coding schemes, etc.}

\rev{The total number of nodes in the network is 
$N=g^2$, where $g$ is the grid size, and simulations have been conducted for 
several values of $g$.}

\subsubsection{\rev{Setup for the random scenario}}

\rev{In the \textit{random scenario}, $N$ nodes are randomly deployed within a 
square of size $l$. The coverage range $r$ of a node is determined as the 
maximum range which allows a success transmission probability above 90\% for a 
packet of size $L$ transmitted with rate $R$ and assuming no fading.} 

\rev{The channel model used in this scenario combines the 
\textit{LogDistancePropagationLossModel} for path loss and the 
\textit{NakagamiPropagationLossModel} to emulate Rayleigh fading. The 
\textit{NistErrorRateModel} validated in \cite{pei2010} was adopted, that takes 
into account the different robustness levels of each modulation and coding 
scheme.}

\rev{The goal of the random scenario is to investigate how the RCFD algorithm 
proposed in this paper would perform in a more realistic ad hoc wireless 
network in 
comparison to the other channel access techniques.}

\subsubsection{\rev{Traffic model and metrics for both scenarios}}

In each node, several applications are installed, one for each node within its 
coverage range, as shown in Fig.~\ref{fig:simulation_scenario} 
for the structured scenario. 
The starting time of each application, $t_s$, is distributed as an exponential 
random variable of parameter $\lambda_s$ truncated after $t_{s,max}$, 
while the stop time coincides with the end of the simulation.
 
An \textit{OnOffApplication} model is adopted where the duration of the ON and 
OFF periods are also exponentially distributed, with mean $T_{ON}$ and 
$T_{OFF}$, respectively. 
During the ON period, the applications generates constant bitrate (CBR) traffic 
with source rate $R_s$. All packets have the same length $L$ and the data 
rate at the physical layer, $R$, is constant.

Network operations have been simulated for a total of $T$ seconds (with the 
initial transient period removed), for different values of the network 
size $N$. Given a certain parameter configuration, 
each simulation has been repeated a total of $N_S$ times and results have been 
averaged. 

We considered two performance metrics, namely the \textit{normalized system 
throughput}, $\Gamma$, and the \textit{average delay}, $\Delta$. 
The normalized system throughput is the ratio of the total number of 
payload bits successfully delivered by all the nodes in the network 
over the simulation time $T$, and the offered traffic $G$.
The offered traffic is given by
\begin{equation}
G=R_s\cdot N_a\cdot 
\frac{T_{ON}}{T_{ON}+T_{OFF}}
\end{equation}
where $N_a$ is the total number of running applications in the network, which 
is a function of the network size $N$ and the coverage radius $r$.

The average delay, on the other hand, is the arithmetic mean of the delay 
experienced by each packet in the network, defined as the time elapsed from the 
instant in which the packet is generated by the application to the 
instant in which the packet is successfully delivered \rev{or 
discarded}.\footnote{\rev{A packet is discarded in three cases: (1) the 
transmission keeps failing after $N_{tx,max}$ transmission attempts; (2) the 
packet transmission queue has exceeded the maximum size $Q_{max}$; (3) the time 
elapsed from the packet generation has exceeded the threshold $\Delta_{max}$.}}

Tab.~\ref{tab:sim_param} reports all the parameters 
adopted in the simulations.

\begin{table}[!tb]
\caption{Simulation parameters}
\label{tab:sim_param}
\centering
\scalebox{0.65}{
\begin{tabular}{l|c|c}
\toprule
{\bfseries Parameter} & {\bfseries Description} & {\bfseries Value}\\
\midrule
$d$ & Distance between two adjacent nodes in the structured scenario & 100~m\\
$l$ & Side of deployment area in the random scenario & 500~m\\
$\lambda_s$ & Parameter of application starting time & $0.5~\textrm{s}^{-1}$\\
$t_{s,max}$ & Maximum application starting time & 5~s\\
$T_{ON}$ & Average time during which each application is ON & 0.1~s\\
$T_{OFF}$ & Average time during which each application is OFF & 0.1~s\\
$R_s$ & Application source rate during the ON period & 1~Mbit/s\\
$T$ & Duration of each simulation & 20~s\\
$N_{tx,max}$ & Maximum number of retransmissions at the MAC layer & 7\\
$Q_{max}$ & Transmission queue size (packets) & 1000\\
$\Delta_{max}$ & Maximum interval after which a packet is discarded & 1~s\\
$L$ & Payload length for packets & $\{200,500,1000\}$~Bytes\\
$R$ & Data rate at the PHY layer & $\{6,18,54\}$~Mbit/s\\
$N_S$ & Number of simulations for each configuration & 10\\
\bottomrule
\end{tabular}}
\end{table}

It is worth noting that the simulation--based results \rev{presented in this 
section} are complementary with respect to those presented in 
Section~\ref{sec:theoretical}, since the \rev{latter were} based on the 
assumption of a single collision domain, whereas in \rev{this section} we allow 
multiple collision domains.

\subsection{Simulation results \rev{for the structured scenario}}
\label{subsec:simulation_results_grid}

In order to provide a comprehensive assessment of the presented protocol, in 
the network simulations we have evaluated its performance in the 
structured scenario for two opposite cases:
\begin{rules}
\item \textit{Long packet transmission time:} in this case large payload
packets ($L=1000$~Bytes) were exchanged at the lowest possible rate provided by 
IEEE 802.11g, namely $R=6$~Mbit/s, resulting in a very long packet 
transmission time.
\item \textit{Short packet transmission time:} in this case small payload 
packets ($L=200$~Bytes) were exchanged at the highest possible rate, namely 
$R=54$~Mbit/s, with a corresponding short packet transmission time.
\end{rules}

In each case, the aforementioned performance metrics for the considered 
MAC algorithms have been evaluated for different values of the grid size 
parameter $g$, ranging from 3 ($N=9$ nodes) to 10 ($N=100$ nodes). 

\begin{figure}[!t]
\includegraphics[width=0.9\columnwidth]{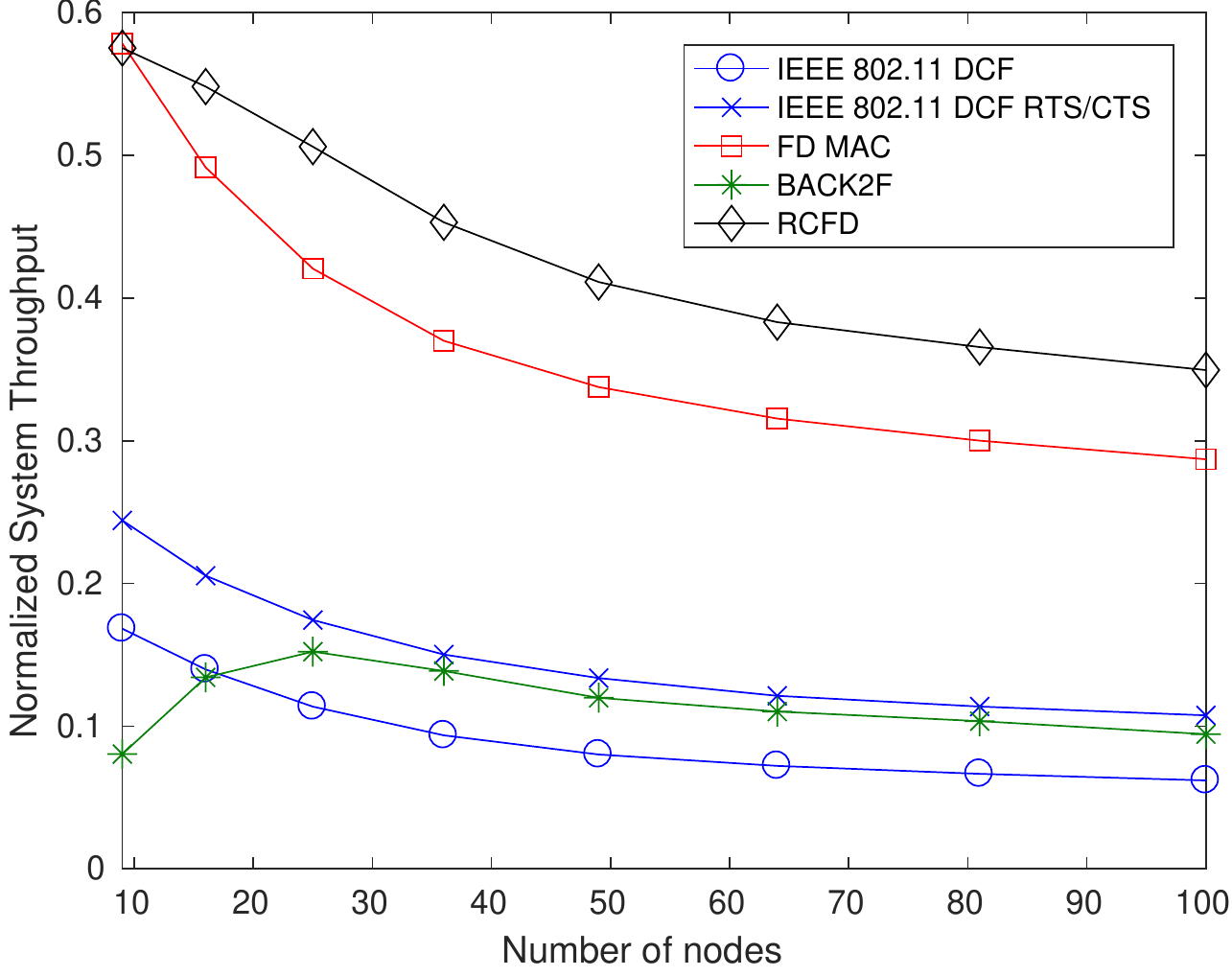}
\caption{Simulated normalized system throughput $\Gamma$ for the 
structured scenario, case I
($R=6$~Mbit/s, $L=1000$~Bytes).}
\label{fig:simulation_high_throughput}
\end{figure}

\begin{figure}[!t]
\includegraphics[width=0.9\columnwidth]{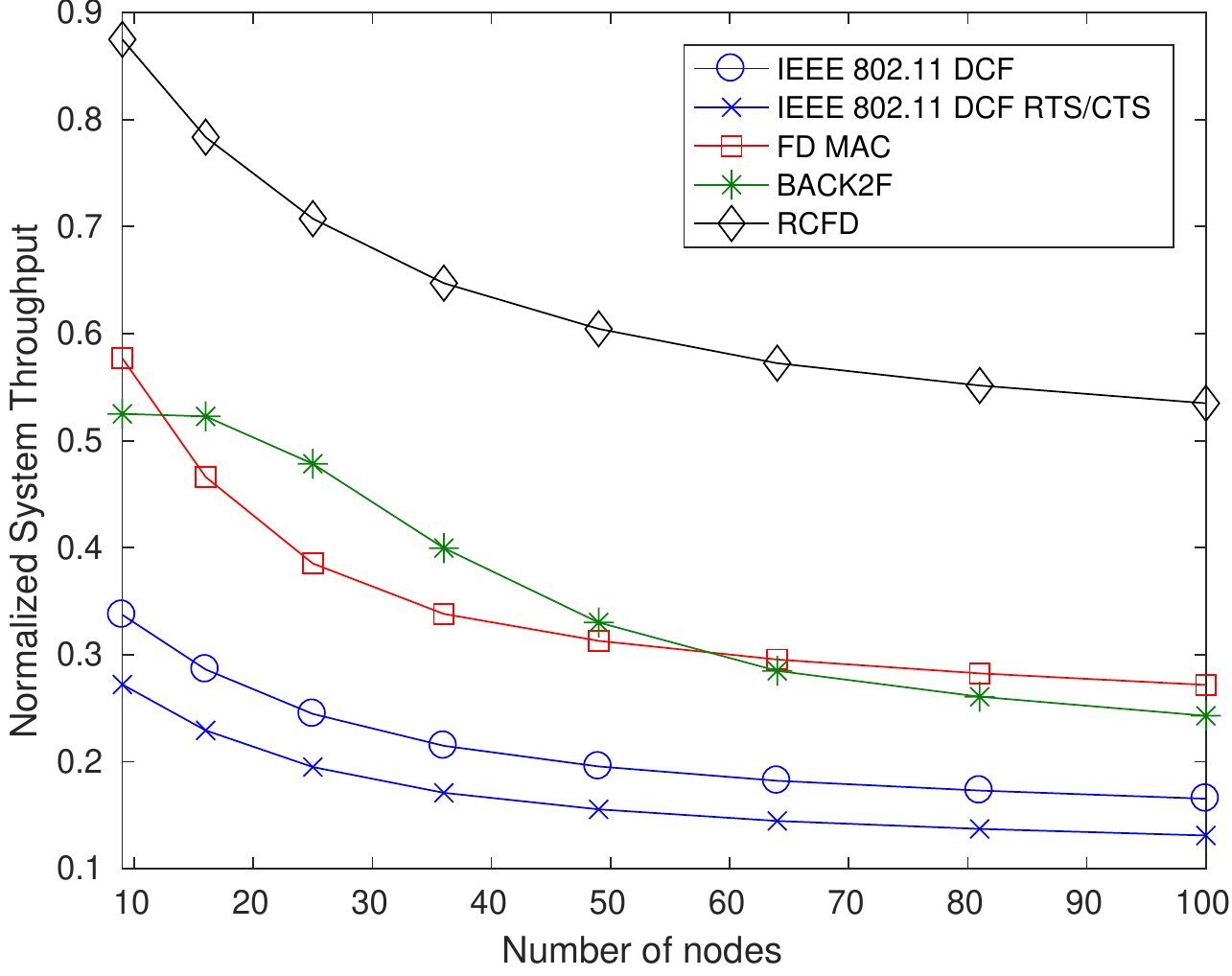}
\caption{Simulated normalized system throughput $\Gamma$ for the 
structured scenario, case II 
($R=54$~Mbit/s, $L=200$~Bytes).}
\label{fig:simulation_low_throughput}
\end{figure}

Fig.~\ref{fig:simulation_high_throughput} shows the normalized system 
throughput $\Gamma$ for case I. \rev{The RCFD strategy outperforms the other 
MAC protocols for any network size. The FD MAC algorithm is able to 
achieve similar performance when the number of nodes is small, but its 
throughput significantly degrades as the network size increases. 
The BACK2F protocol presents a significantly lower $\Gamma$, due to its 
difficulties in handling multiple collision domains. Finally, the IEEE 
802.11 strategies based on time domain channel contention perform poorly.}

The same metric $\Gamma$ is reported in 
Fig.~\ref{fig:simulation_low_throughput} for the second case. Again,
\rev{RCFD performs much better than all other strategies. It can be observed, 
in particular, that the 
schemes relying upon the exchange of RTS/CTS frames (FD MAC and IEEE 802.11) 
perform much worse than in the previous case, since these frames represent a 
significant overhead, given the lower time needed for the actual transmission 
of data frames. BACK2F, which instead relies on frequency domain contention as 
RCFD, performs much better than in the previous case, reaching similar 
performance as FD MAC, despite not being a full--duplex MAC protocol.}

It is worth noticing that the normalized throughput values are 
higher in Fig.~\ref{fig:simulation_low_throughput} with respect to 
Fig.~\ref{fig:simulation_high_throughput}. Indeed, the higher PHY rate allows 
to exchange an increased amount of data in the same time. 

\begin{figure}[!t]
\includegraphics[width=0.9\columnwidth]{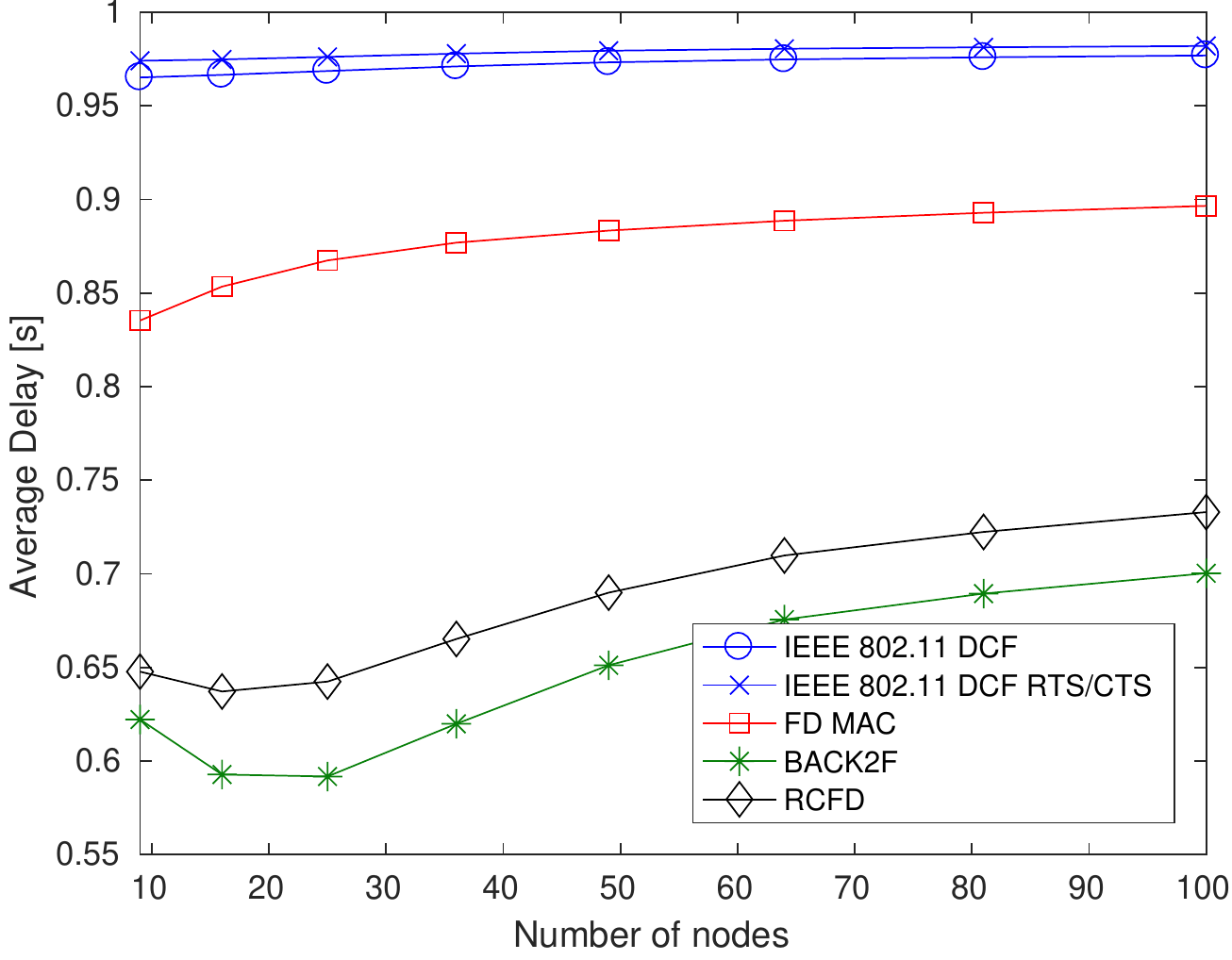}
\caption{Simulated average delay $\Delta$ for the structured scenario, 
case II ($R=54$~Mbit/s, 
$L=200$~Bytes).}
\label{fig:simulation_low_delay}
\end{figure}

The average delay $\Delta$ simulated in case II for all the MAC protocols 
is shown in Fig.~\ref{fig:simulation_low_delay}. \rev{The strategies that 
include frequency domain channel contention strongly outperform those based on 
a time domain approach. 
In particular, BACK2F slightly outperforms the RCFD strategy, mostly due 
to the lower number of contention rounds in the frequency domain (2 against 3). 
Similar results are achieved for case I, not reported here for space 
constraints.} 

\subsection{\rev{Simulation results for the random scenario}}
\label{subsec:simulation_results_random}

\rev{In the random scenario, the performance of the considered MAC algorithms 
have been evaluated for different network size values, ranging from $N=10$ to 
$N=50$ nodes. The payload size has been fixed to $L=500$~Bytes and the PHY 
layer transmission rate to $R=18$~Mbps, providing an intermediate case between 
the two extremes analyzed in the structured scenario. Under this configuration, 
the coverage radius of each node was set to $r=60$~m in order to provide 90\% 
transmission success probability.}

\begin{figure}[!t]
\includegraphics[width=0.9\columnwidth]{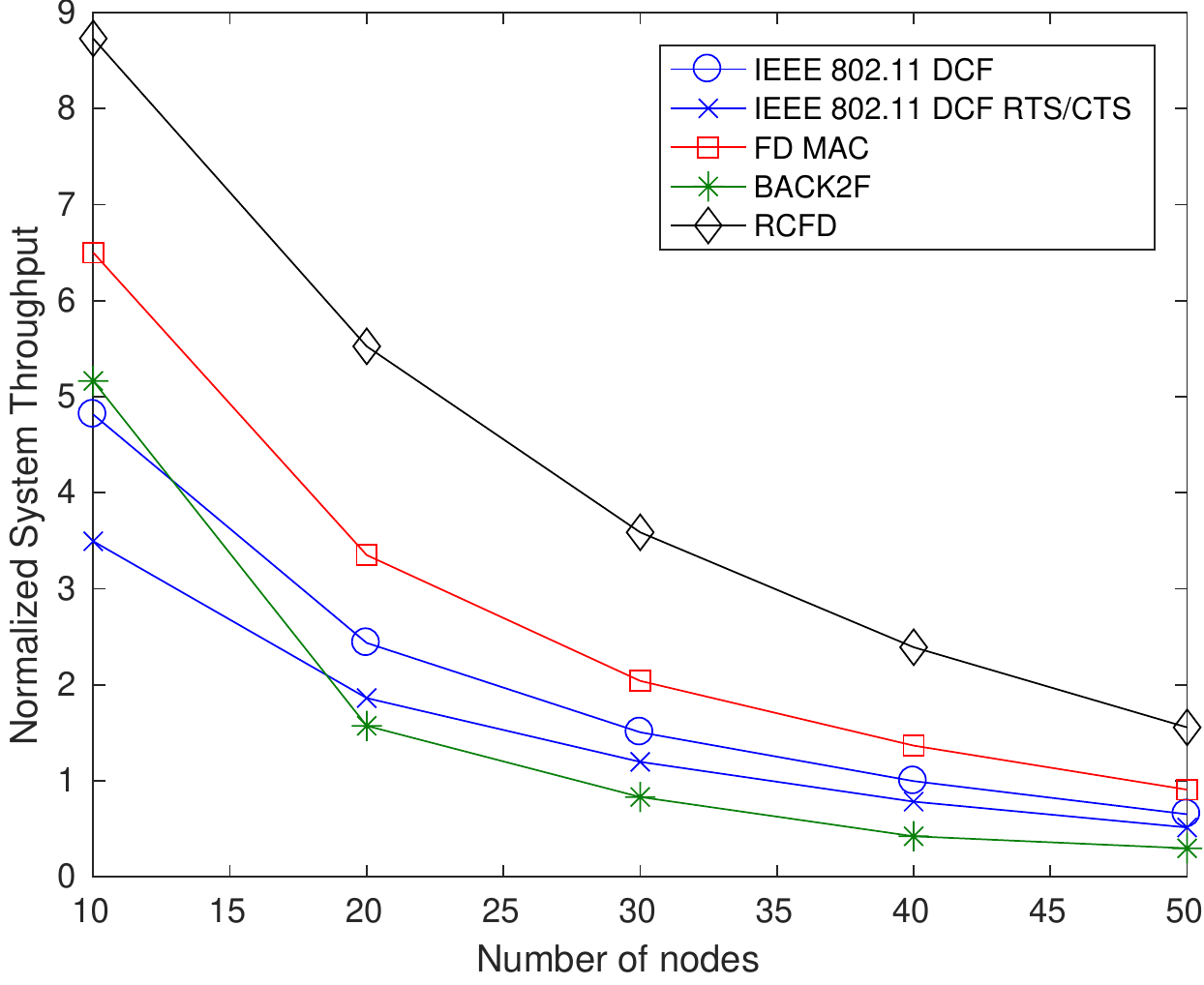}
\caption{\rev{Simulated normalized system throughput $\Gamma$ for the random 
scenario ($R=18$~Mbit/s, $L=1000$~Bytes).}}
\label{fig:simulation_random_throughput}
\end{figure}

\rev{Fig.~\ref{fig:simulation_random_throughput} shows the normalized system 
throughput $\Gamma$ for the different MAC algorithms. Also in this case, RCFD 
is able to significantly outperform all the other schemes. As in case I of the 
structured scenario, FD MAC provides the closest performance, while the 
throughput of the BACK2F algorithm suffers from the presence of multiple 
collision domains and significantly degrades with the network size.}

\begin{figure}[!t]
\includegraphics[width=0.9\columnwidth]{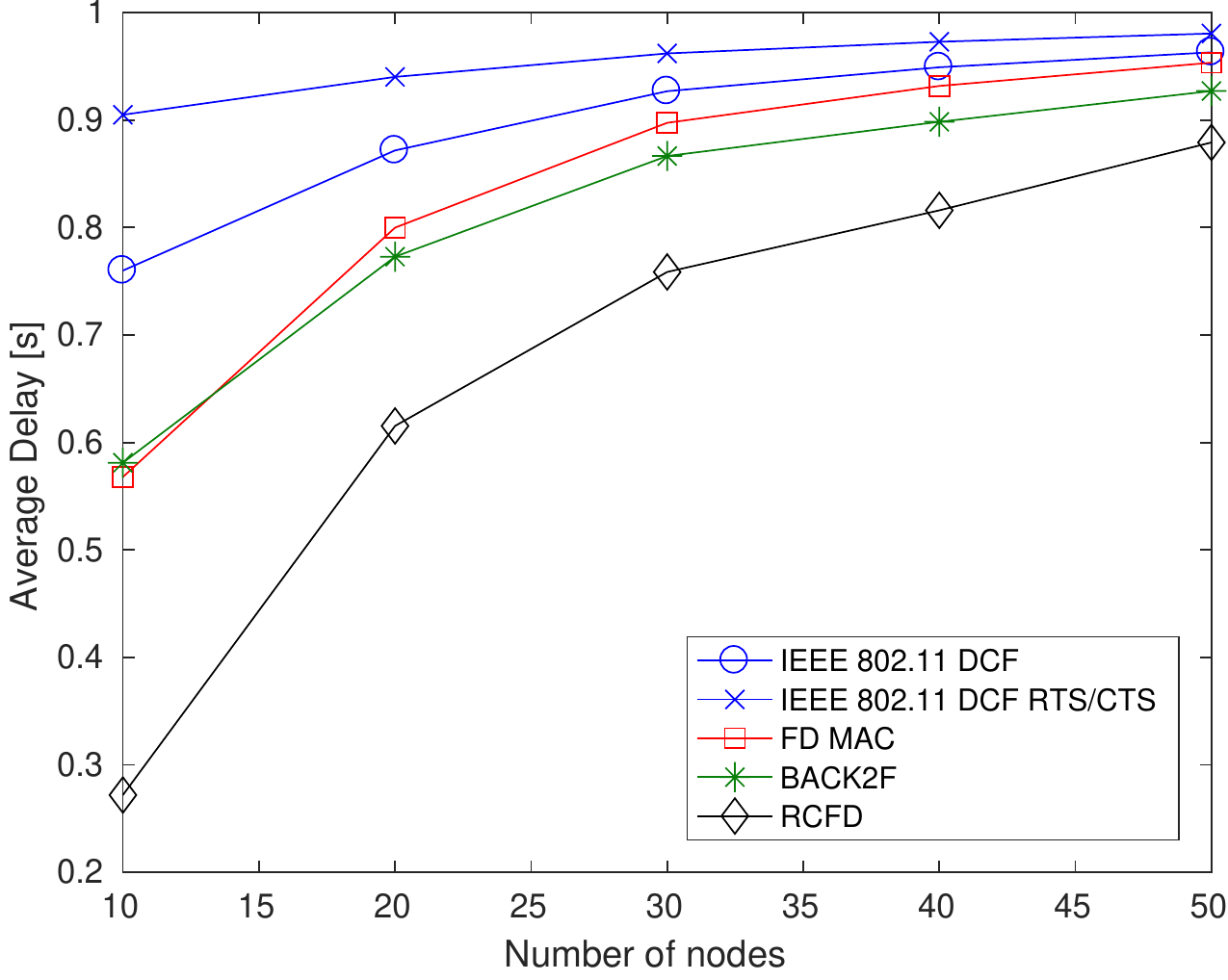}
\caption{\rev{Simulated average delay $\Delta$ for the random 
scenario ($R=18$~Mbit/s, $L=1000$~Bytes).}}
\label{fig:simulation_random_delay}
\end{figure}

\rev{The average delay $\Delta$ for the random scenario is reported in 
Fig.~\ref{fig:simulation_random_delay}. Again, RCFD significantly outperforms 
all the other schemes, confirming that this strategy represents a very 
interesting opportunity for real--world applications. 
Among the other algorithms, BACK2F emerges as the one able to guarantee the 
lowest delay, thanks to the channel contention in the frequency domain.}

\begin{figure}[!t]
\includegraphics[width=0.9\columnwidth]{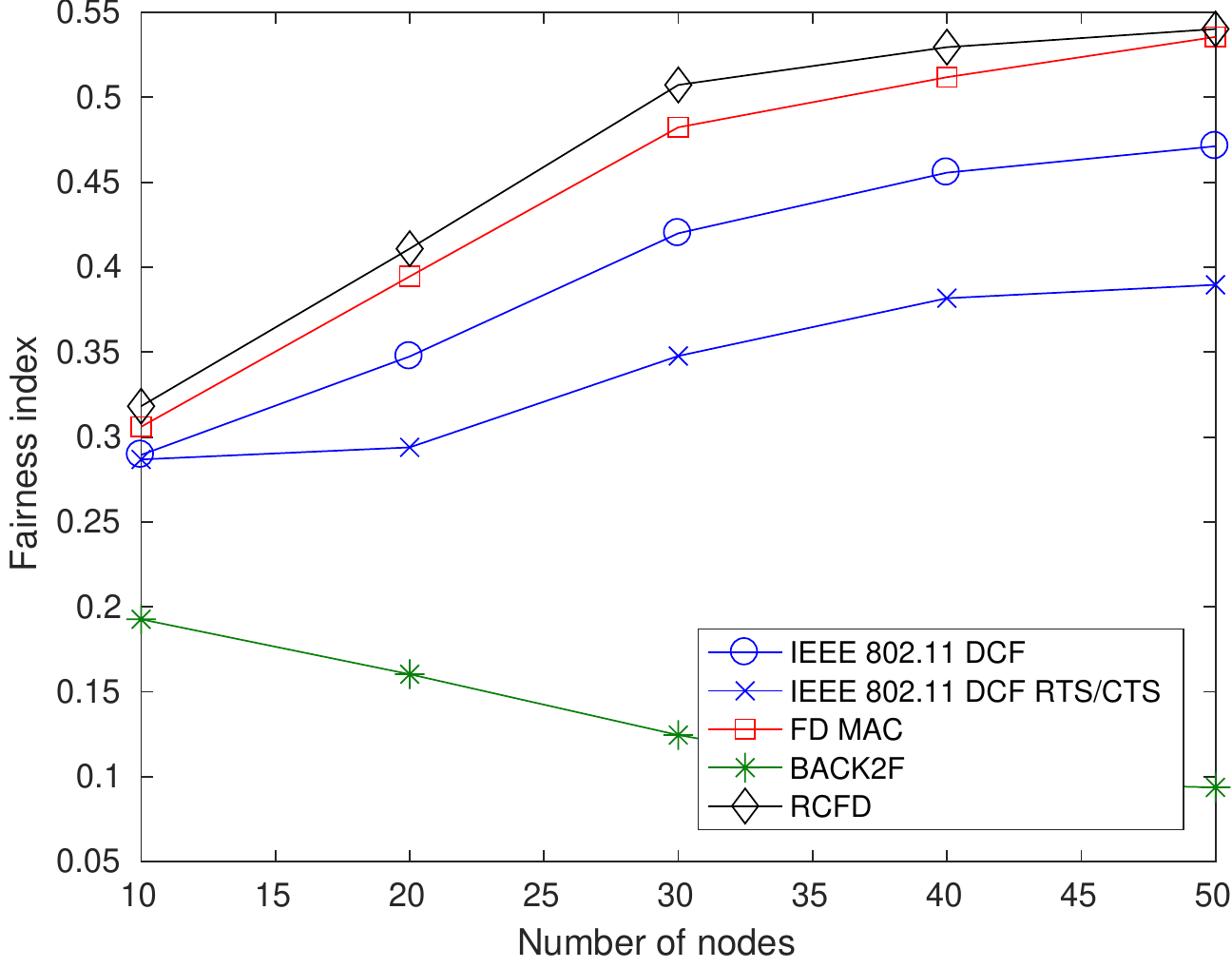}
\caption{\rev{Simulated fairness index $J$ for the random 
scenario ($R=18$~Mbit/s, $L=1000$~Bytes).}}
\label{fig:simulation_random_fairness}
\end{figure}

\rev{In order to provide a final insight, the fairness of the compared MAC 
protocols is reported in Fig.~\ref{fig:simulation_random_fairness} for the 
random scenario, measured in terms of Jain's fairness index \cite{jain1984}, 
defined as
\begin{equation}
\mathcal{J}\left(p_1,\dots,p_N\right)=\frac{\left(\sum\limits_{i=1}^N 
p_i\right)^2}{N\cdot\sum\limits_{i=1}^N p_i^2}
\end{equation}
where $p_i$ is the number of packets successfully received by node $n_i$. It 
can be observed that, also in terms of fairness, RCFD outperforms all other 
protocols.}

\section{Conclusions}
\label{sec:conclusions}

The currently employed channel access schemes for wireless networks 
present several issues and relatively low performance. The introduction of 
full--duplex wireless communication can lead to increased performance but also 
poses additional challenges to transmission scheduling, and no standard MAC 
protocol has emerged so far as the best solution for FD wireless networks. In 
this paper we proposed RCFD, a full--duplex MAC protocol based on a 
time--frequency channel access procedure. We showed through theoretical 
analyses and network simulations that this strategy provides excellent 
performance in terms of both throughput and packet transmission delay, also in 
the case of dense networks, compared to other standard and state--of--the--art 
MAC layer schemes.

A natural extension of this work is the experimental assessment of the 
proposed MAC layer protocol on devices capable of FD operations and able to 
transmit OFDM symbols using only some specific subcarriers. The optimizations 
to the presented protocol suggested in Section~\ref{subsec:optimization} 
could lead to even higher performance gains with respect to other MAC layer 
strategies. \rev{Also, the protocol can be extended to account for asymmetric 
full--duplex in addition to bidirectional one. Finally, the performance of RCFD 
can be tested in the more realistic cases of imperfect FD cancellation and/or 
mixed networks with both HD and FD nodes.}

\bibliographystyle{IEEEtran}
\bibliography{IEEEfull,biblio_fdw}

\setcounter{equation}{28}

\appendices
\section{Derivation of transition probabilities for the Markov chain used in 
the BACK2F analysis}
\label{sec:appendix}

The transition probabilities for the Markov chain defined in 
Section~4.2 are of the form
\begin{equation}
\footnotesize
p_{i,a,j\vert k,b,l}=P\left\{x(t)=i,c(t)=a,y(t)=j\,\vert\, 
x(t-1)=k,c(t-1)=b,y(t-1)=l\right\}
\end{equation}
Through some computations, $p_{i,a,j\vert k,b,l}$ can be factorized in three 
terms
\begin{equation}
\label{eq:factorized}
p_{i,a,j\vert k,b,l}=p_{j\vert i,a,k,b,l}\cdot p_{i\vert a,k,b,l}\cdot 
p_{a\vert k,b,l}
\end{equation}
where 
{\footnotesize
\begin{align}
p_{j\vert i,a,k,b,l}&=P\{y(t)=j\,\vert\, 
x(t)=i,c(t)=a,x(t-1)=k,c(t-1)=b,y(t-1)=l\}\\
p_{i\vert a,k,b,l}&=P\{x(t)=i\vert c(t)=a,x(t-1)=k,c(t-1)=b,y(t-1)=l\}\\
p_{a\vert k,b,l}&=P\{c(t)=a\vert x(t-1)=k,c(t-1)=b,y(t-1)=l\}
\end{align}}
In the remainder of this appendix, exact expressions for these three terms are 
derived for all possible values of the parameters $i,a,j,k,b,l$, according to 
the structure of the BACK2F algorithm, explained in detail in [22] 
and reported in Algorithm~\ref{alg:back2f} for convenience.

\begin{algorithm}
  \caption{BACK2F channel access algorithm [22].}
  \label{alg:back2f}
\begin{algorithmic}[1] 
\begin{small}
\Procedure{BACK2F}{\textit{packet}}
    \State $\mathtt{myback} \gets \mathit{rnd} \left[0,S-1\right]$
    \State wait for $T_{difs}$
    \If{channel is busy}
    	\State \textbf{goto} line 2
    \Else
    	\State transmit on SC \texttt{myback} in round 1
    	\State $\mathtt{minback} \gets$ lowest--frequency SC with signal
    	\State $\mathtt{myback} \gets \mathtt{myback} - \mathtt{minback}$
    	\If{$\mathtt{myback}>0$} \Comment{Lost round 1}
    		\State \textbf{goto} line 2
    	\Else
    		\State $\mathtt{myback2} \gets \mathit{rnd} \left[0,S-1\right]$
    		\State transmit on SC \texttt{myback2} in round 2
    		\State $\mathtt{minback2} \gets$ lowest--frequency SC with signal	
    		\If{$\mathtt{myback2}=\mathtt{minback2}$} \Comment{Won round 2}
    		    \State transmit \textit{packet}
    		    \State \textbf{goto} line 1
    		\Else	\Comment{Lost round 2}
    			\State \textbf{goto} line 2
    		\EndIf 
    	\EndIf 
    \EndIf   
\EndProcedure
\end{small}
\end{algorithmic}
\end{algorithm}

\subsection{Derivation of $\mathbf{p_{j\vert i,a,k,b,l}}$}

It can first be observed that
{\footnotesize
\begin{align}
\label{eq:ana_first}
p_{j\vert i,a,k,b,l} &= P\left\{y(t)=j\,\vert\, 
x(t)=i,c(t)=a,x(t-1)=k,c(t-1)=b,y(t-1)=l\right\}\nonumber\\
&= P\left\{y(t)=j\,\vert\, x(t)=i\right\}
\end{align}} %
since the number of winning nodes at the second round only depends on the 
number of 
nodes that have won the first round in the same time slot. We can further state 
that \eqref{eq:ana_first} is meaningful only for $j\leq i$, which leaves 
only the two following scenarios.

\begin{scenario}[leftmargin=5mm]
\item $j=i$

In this case we have $i$ nodes randomly choosing among $S$ subcarriers. The 
probability that they all pick the same one is given by $1/S^{i-1}$.
\item $j<i$

The probability that $j$ nodes out of $i$ pick the same SC $c$ and that all the 
other nodes pick SCs with higher frequency than $c$ is given by
\begin{equation}
{{i}\choose{j}}\left(\frac{1}{S}\right)^j\left(1-\frac{c+1}{S}\right)^{i-j}
\end{equation}

This probability has to be summed over all possible SCs except the last one 
(which would result in all the nodes picking the same one, i.e., $j=i$)
\begin{equation}
\sum\limits_{c=0}^{S-2}{{i}\choose{j}}\left(\frac{1}{S}\right)^j\left(1-\frac{c+1}{S}\right)^{i-j}
\end{equation}
\end{scenario}

Summing up all the scenarios, we obtain the following expression for $p_{j\vert 
i,a,k,b,l}$:
\begin{equation}
p_{j\vert i,a,k,b,l}=
\begin{cases}
\sum\limits_{c=0}^{S-2}{{i}\choose{j}}\left(\frac{1}{S}\right)^j\left(1-\frac{c+1}{S}\right)^{i-j}
 & \text{if } j<i\\
\frac{1}{S^{i-1}} & \text{if } j=i\\
0 & \text{otherwise}
\end{cases}
\end{equation}

\subsection{Derivation of $\mathbf{p_{i\vert a,k,b,l}}$}
\label{subsec:second_term}

To compute this second term we have to derive the probability that exactly $i$ 
nodes win the first round at time slot $t$ given that SC $a$ is the 
lowest--frequency one and that, in the previous time slot, $k$ nodes won the 
first round (with SC $b$) and $l$ nodes won the second round. Again, we split 
the problem in multiple scenarios.

\begin{scenario}[leftmargin=5mm]
\item $k\neq l$

In this scenario, at the end of time slot $t-1$ we have the following groups of 
nodes:
\begin{groups}
\item $N-k$ nodes that have lost round 1 and, hence, have $\mathtt{myback}>0$ 
(line 10 in Algorithm~\ref{alg:back2f}).
\item $k-l\neq 0$ nodes that have lost round 2 and, hence, have 
$\mathtt{myback}=0$ 
(line 12 in Algorithm~\ref{alg:back2f}).
\item $l$ nodes that have won round 2 and, after transmitting, have 
$\mathtt{myback}$ randomly distributed between 0 and S-1 (lines 18 and 2 in 
Algorithm~\ref{alg:back2f}).
\end{groups}

Therefore, the following observations can be made:
\begin{itemize}
\item The lowest--frequency SC at time $t$ is 0 (chosen by at least the 
nodes of group B), hence $p_{i\vert a,k,b,l}$ is always 0 when $a\neq 0$. 
\item There are at least $k-l$ nodes (group B) that have $\mathtt{myback}=0$ 
and win 
round 1, hence, $i\geq k-l$. 
\item The maximum number of first round winners is $k$, since $N-k$ nodes 
(group A) have $\mathtt{myback}>0$, hence, $i<k$. 

The probability that $m$ of the $l$ nodes of group C pick 0 as a SC and hence 
win round 1 at time slot $t$ is
\begin{equation}
\label{eq:ana_second_scenario1}
{{l}\choose{m}}\left(\frac{1}{S}\right)^m\left(1-\frac{1}{S}\right)^{l-m}
\end{equation}

and the corresponding number $i$ of first--round winners is $i=k-l+m$, hence, 
we obtain $p_{i\vert a,k,b,l}$ for the case of $a=0$ and $k\neq l$ by replacing 
$m$ in \eqref{eq:ana_second_scenario1} with $i-k+l$.
\end{itemize}

\begin{figure*}[!t]
\setcounter{MYtempeqncnt}{\value{equation}}
\setcounter{equation}{43}
\begin{equation}
p_{i\vert a,k,b,l}=
\begin{cases}
{{l}\choose{i-k+l}}\left(\frac{1}{S}\right)^{i-k+l}\left(1-\frac{1}{S}\right)^{k-i}
 & \text{if } k\neq l,a=0,k-l\leq i\leq k\\
\frac{{{k}\choose{i}}\left(\frac{1}{S}\right)^i\left(1-\frac{1}{S}\right)^{k-i}}{1-\left(1-\frac{1}{S}\right)^k}
 & \text{if } k=l\neq N, a=0, i\leq k\\
{{k}\choose{i-N+k}}\left(\frac{1}{b+1}\right)^{i-N+k}\left(1-\frac{1}{b+1}\right)^{N-i}
 & \text{if } k\!=\!l\!\neq\! N, a=\!S\!-\!b-\!1, i\!\geq\!N\!-\!k\\
\frac{\left[{{N-k}\choose{n}}\left(\frac{1}{S-b-a}\right)^n\left(1-\frac{1}{S-b-a}\right)^{N-k-n}\right]\cdot
\left[{{k}\choose{i-n}}\left(\frac{1}{S-a}\right)^{i-n}\left(1-\frac{1}{S-a}\right)^{k-i+n}\right]}
{1-\left(1-\frac{1}{S-a}\right)^k\left(1-\frac{1}{S-b-a}\right)^{N-k}}
 & \text{if } k=l\neq N, 0<a<S-b-1\\
\frac{{{N}\choose{i}}\left(\frac{1}{S-a}\right)^i\left(1-\frac{1}{S-a}\right)^{N-i}}
{1-\left(1-\frac{1}{S-a}\right)^N}
 &  \text{if } k=l=N\\
0 & \text{otherwise}
\end{cases}
\label{eq:ana_second_sumup}
\end{equation}
\setcounter{equation}{\value{MYtempeqncnt}}
\hrulefill
\end{figure*}

\item $k=l\neq N$

In this scenario, at the end of time slot $t-1$ we have the following group of 
nodes:
\begin{groups}
\item $N-k$ nodes have lost round 1 and, hence, will have $\mathtt{myback}>0$. 
The maximum value of \texttt{myback} for this node is $S-b-1$, according to 
line 9 in Algorithm~\ref{alg:back2f} and taking into account that 
$\mathtt{minback}=b$ (lowest--frequency SC at round 1 in time slot $t-1$).
\item $k$ nodes have won round 2 and, after transmitting, will have 
$\mathtt{myback}$ randomly varying between 0 and S-1 (lines 18 and 2 in 
Algorithm~\ref{alg:back2f}).
\end{groups}

The case $a=0$ is trivial, since the maximum number of first round winners is 
$k$ (analogously to scenario I) and the probability that $i$ nodes out of $k$ 
(group B) select $\mathtt{myback}=0$ (given that there is at least one node 
that 
selects it) is
\begin{equation}
\label{eq:ana_second_scenario2a}
\frac{{{k}\choose{i}}\left(\frac{1}{S}\right)^i\left(1-\frac{1}{S}\right)^{k-i}}{1-\left(1-\frac{1}{S}\right)^k}
\end{equation}

Another trivial case is $a=S-b-1$: in this situation, the $N-k$ nodes of group 
A all win the first round at $t$ (hence $i\geq N-k$) and the probability that 
$m$ nodes out of the remaining $k$ (group B) select 
SC $S-b-1$ is
\begin{equation}
\label{eq:ana_second_scenario2b}
{{k}\choose{m}}\left(\frac{1}{b+1}\right)^m\left(1-\frac{1}{b+1}\right)^{k-m}
\end{equation}
with $i=N-k+m$.

The case of $0<a<S-b-1$, instead, is non--trivial, since the nodes from both 
groups can select $a$ as a SC. In detail, the probability that $n$ nodes from 
group A and $i-n$ nodes from group B select SC $a$ (given that at least one 
node selects it) is given by
\begin{equation}
\label{eq:ana_second_scenario2c}
\frac{\left[{{N-k}\choose{n}}\left(\frac{1}{S-b-a}\right)^n\left(1-\frac{1}{S-b-a}\right)^{N-k-n}\right]\cdot
\left[{{k}\choose{i-n}}\left(\frac{1}{S-a}\right)^{i-n}\left(1-\frac{1}{S-a}\right)^{k-i+n}\right]}
{1-\left(1-\frac{1}{S-a}\right)^k\left(1-\frac{1}{S-b-a}\right)^{N-k}}
\end{equation}

The expression in \eqref{eq:ana_second_scenario2c} has to be summed for all 
possible values of $n$, taking into account that $0\leq n\leq i$ by definition, 
and also $n\leq N-k$ and $i-n\leq k$. Therefore, the probability that $i$ nodes 
win the first round at time $t$ when $k=l\neq N$ and $0<a<S-b-1$ is
\begin{equation}
\footnotesize
\label{eq:ana_second_scenario2c_bis}
\frac{\sum\limits_{n=\max(i-k,0)}^{min(N-k,i)}\left[{{N-k}\choose{n}}\left(\frac{1}{S-b-a}\right)^n\left(1-\frac{1}{S-b-a}\right)^{N-k-n}\right]\cdot
\left[{{k}\choose{i-n}}\left(\frac{1}{S-a}\right)^{i-n}\left(1-\frac{1}{S-a}\right)^{k-i+n}\right]}
{1-\left(1-\frac{1}{S-a}\right)^k\left(1-\frac{1}{S-b-a}\right)^{N-k}}
\end{equation}
  
\item $k=l=N$

In this scenario there is only one group of $N$ nodes, which have all won the 
second round in time slot $t-1$ and hence can select \texttt{myback} in the 
range $[a,S-1]$. The probability that exactly $i$ nodes select 
$\mathtt{myback}=a$ (given that at least one selects it) is given by
\begin{equation}
\label{eq:ana_third}
\frac{{{N}\choose{i}}\left(\frac{1}{S-a}\right)^i\left(1-\frac{1}{S-a}\right)^{N-i}}
{1-\left(1-\frac{1}{S-a}\right)^N}
\end{equation}
\end{scenario}

Summing up all the scenarios, we obtain the expression for $p_{i\vert 
a,k,b,l}$ reported in \eqref{eq:ana_second_sumup}.

\subsection{Derivation of $\mathbf{p_{a\vert k,b,l}}$}

To compute the third and last term, we have to derive the probability that SC 
$a$ is the lowest--frequency one at the first contention round during time slot 
$t$, given that, in the previous time slot, $k$ nodes won the first round (with 
SC $b$) and $l$ nodes won the second round. We consider the same three 
scenarios as in Sec.~\ref{subsec:second_term}, thus, for each scenario, we 
refer to the same groups of nodes.

\begin{scenario}[leftmargin=5mm]
\item $k\neq l$

As explained in Sec.~\ref{subsec:second_term}, the only possible value for $a$ 
in this scenario is 0, hence $p_{0\vert k,b,l}=1$ and 0 otherwise.

\item $k=l\neq N$

As explained in Sec.~\ref{subsec:second_term}, the case $a=0$ is trivial, since 
it can only happen for the $k$ nodes of group B and it happens with probability
\begin{equation}
\label{eq:ana_third_scenario2a}
1-\left(1-\frac{1}{S}\right)^k
\end{equation}

\begin{figure*}[!t]
\setcounter{MYtempeqncnt}{\value{equation}}
\setcounter{equation}{51}
\begin{equation}
p_{a\vert k,b,l}=
\begin{cases}
1
 & \text{if } k\neq l,a=0\\
1-\left(1-\frac{1}{S}\right)^k
 & \text{if } k=l\neq N, a=0\\
\left(1-\frac{a}{S}\right)^k\left(1-\frac{a-1}{S-b-1}\right)^{N-k} - 
\left(1-\frac{a+1}{S}\right)^k\left(1-\frac{a}{S-b-1}\right)^{N-k}
 & \text{if } k\!=\!l\!\neq\! N, 0\!<\!a\!\leq \!S\!-\!b\!-\!1\\
\left(1-\frac{a}{S}\right)^N-\left(1-\frac{a+1}{S}\right)^N
 &  \text{if } k=l=N\\
0 & \text{otherwise}
\end{cases}
\label{eq:ana_third_sumup}
\end{equation}
\setcounter{equation}{\value{MYtempeqncnt}}
\hrulefill
\end{figure*}

The case $a>0$, instead, is non--trivial. We also have that $a\leq S-b-1$, as 
discussed in Sec.~\ref{subsec:second_term}. Let us indicate with a random 
variable $X_i,i=1,\dots,S-b-1$ the number of nodes in group A that select SC 
$i$ at the first round and with $Y_j,j=0,\dots,S-1$ the number of nodes in 
group B that select SC $j$ at the first round. Both these groups of random 
variables follow a multinomial distribution with constant probabilities 
$p_i=\frac{1}{S-b-1}$ for the first group and $p_j=\frac{1}{S}$ for the second 
group.
The probability that $a$ is the lowest--frequency SC is expressed as 
\begin{equation}
\label{eq:ana_third_scenario2a_all}
p_{a\vert k,b,l} = p_{X,a} \cdot p_{Y,a} + p_{X,a}\cdot p_{Y,\bar{a}} + 
p_{X,\bar{a}}\cdot p_{Y,a}  
\end{equation}
where:
\begin{align}
\nonumber
p_{X,a}&=P\left\{X_1=0,\dots,X_{a-1}=0,X_a\neq 0\right\}\\
\nonumber
&=P\left\{X_1=0,\dots,X_{a-1}=0\right\}-P\left\{X_1=0,\dots,X_a=0\right\}\\
\label{eq:ana_third_scenario2a_1}
&= \left(1-\frac{a-1}{S-b-1}\right)^{N-k}-\left(1-\frac{a}{S-b-a}\right)^{N-k}\\
\nonumber
p_{X,\bar{a}}&=P\left\{X_1=0,\dots,X_a = 0\right\}\\
\label{eq:ana_third_scenario2a_2}
&= \left(1-\frac{a}{S-b-a}\right)^{N-k}\\
\nonumber
p_{Y,a}&=P\left\{Y_0=0,\dots,Y_{a-1}=0,Y_a\neq 0\right\}\\
\nonumber
&=P\left\{Y_0=0,\dots,Y_{a-1}=0\right\}-P\left\{Y_1=0,\dots,Y_a=0\right\}\\
\label{eq:ana_third_scenario2a_3}
&= \left(1-\frac{a}{S}\right)^k-\left(1-\frac{a+1}{S}\right)^k\\
\nonumber
p_{Y,\bar{a}}&=P\left\{Y_0=0,\dots,Y_a = 0\right\}\\
\label{eq:ana_third_scenario2a_4}
&= \left(1-\frac{a+1}{S}\right)^k
\end{align}

The expression for $p_{a\vert k,b,l}$ when $k=l\neq N$ and $a>0$ can hence be 
obtained by inserting \eqref{eq:ana_third_scenario2a_1}, 
(\ref{eq:ana_third_scenario2a_2}), (\ref{eq:ana_third_scenario2a_3}) and 
(\ref{eq:ana_third_scenario2a_4}) in \eqref{eq:ana_third_scenario2a_all} 
\begin{equation}
\label{eq:ana_third_scenario2a_final}
\left(1-\frac{a}{S}\right)^k\left(1-\frac{a-1}{S-b-1}\right)^{N-k} - 
\left(1-\frac{a+1}{S}\right)^k\left(1-\frac{a}{S-b-1}\right)^{N-k}
\end{equation}
 
\item $k=l=N$

In this scenario there is only one group of $N$ nodes, which have all won the 
second round at time slot $t-1$ and hence can select \texttt{myback} in the 
range $[0,S-1]$. Let us indicate with a random variable $Z_i,i=0,\dots,S-1$ the 
number of nodes that select SC $i$ at the first round, multinomially 
distributed with constant probability $p_i=\frac{1}{S}$. The probability that 
$a$ is the lowest--frequency SC is expressed as 
\begin{align}
\nonumber 
p_{a\vert k,b,l} &= P\left\{Z_0=0,\dots,Z_{a-1}=0,Z_{a}\neq 0 \right\}\\  
\nonumber
&= P\left\{Z_0=0,\dots,Z_{a-1}=0\right\}-P\left\{Z_0=0,\dots,Z_a=0\right\}\\
\label{eq:ana_third_scenario3}
&= \left(1-\frac{a}{S}\right)^N-\left(1-\frac{a+1}{S}\right)^N
\end{align}

\end{scenario}

Summing up all the scenarios, we obtain the expression for $p_{a\vert 
k,b,l}$ reported in \eqref{eq:ana_third_sumup}.

\end{document}